\documentclass[journal]{IEEEtran}
\ifCLASSINFOpdf
  \usepackage[pdftex]{graphicx}
  \graphicspath{{../pdf/}{../jpeg/}}
  \DeclareGraphicsExtensions{.pdf,.jpeg,.png}
\else
  \usepackage[dvips]{graphicx}
  \graphicspath{{../eps/}}
  \DeclareGraphicsExtensions{.eps}
\fi
\ifCLASSOPTIONcompsoc
\usepackage[caption=false, font=normalsize, labelfont=sf, textfont=sf]{subfig}
\else
\usepackage[caption=false, font=footnotesize]{subfig}

\usepackage{amsmath}
\usepackage{esint}
\usepackage{amssymb}
\usepackage{breqn}
\usepackage{cite}
\usepackage{color}
\usepackage{booktabs}

\ifCLASSOPTIONcompsoc
 \usepackage[caption=false,font=normalsize,labelfont=sf,textfont=sf]{subfig}
\else
 \usepackage[caption=false,font=footnotesize]{subfig}
\fi
\newcounter{MYtempeqncnt}

\begin{document}
%
\title{Convolutional Beamspace Beamforming for Low-Complexity Far-Field and Near-Field MU-MIMO Communications}
%
%
%

\author{Chao~Feng, Huizhi~Wang, and 
        Yong~Zeng,~\IEEEmembership{Senior~Member,~IEEE}
\thanks{The authors are with the National Mobile Communications Research Laboratory, 
Southeast University, Nanjing 210096, China 
(e-mail: \{chao\_feng, wanghuizhi, yong\_zeng\}@seu.edu.cn).
Yong Zeng is also
with Purple Mountain Laboratories, Nanjing 211111, China.
\textit{(Corresponding author: Yong Zeng.)}}
}
\maketitle

\begin{abstract}
Inter-user interference (IUI) mitigation has been an essential issue for multi-user multiple-input multiple-output (MU-MIMO) communications. The commonly used linear processing schemes include the maximum-ratio combining (MRC), zero-forcing (ZF) and minimum mean squared
error (MMSE) beamforming, which may result in the unfavorable performance or complexity as the antenna number grows. In this paper, we introduce a low-complexity linear beamforming solution for the IUI mitigation by using the convolutional beamspace (CBS) technique. Specifically, the dimension of channel matrix can be significantly reduced via the CBS preprocessing, thanks to its beamspace and spatial filtering effects. 
However, existing methods of the spatial filter design
mainly benefit from the Vandermonde structure of channel matrix, which only holds for the far-field scenario with the uniform plane wave (UPW) model. As the antenna size increases, this characteristic may vanish in the near-field region of the array, where the uniform spherical wave (USW) propagation becomes dominant. To gain useful insights, we first investigate the beamforming design and performance analysis of the CBS-based beamforming based on the UPW model. Our results unveil that the proposed CBS-based MMSE beamforming is able to achieve a near-optimal performance but demands remarkably lower complexity than classical ZF and MMSE schemes. Furthermore, our analysis is also extended to the near-field case. To this end, a novel optimization-based CBS approach is proposed for preserving spatial filtering effects, thus rendering the compatibility of the CBS-based beamforming. Finally, numerical results are provided to demonstrate the effectiveness of our proposed CBS-based beamforming method.

\end{abstract}


%
\IEEEpeerreviewmaketitle


\section{Introduction}

Multi-user multiple-input multiple-output (MU-MIMO) is a key technique to significantly enhance the system throughput of wireless communication networks, thanks to its diversity and multiplexing gains brought by multiple antennas and multiple users \cite{tse2004diversity,spencer2004introduction,gesbert2007shifting,castaneda2016overview}.
To fully reap potentials of MU-MIMO systems, different users may share the same channel resource, leading to the inter-user interference (IUI). In particular, severe IUI may largely degrade the communication performance. Thus, one central problem for MU-MIMO is to mitigate the IUI efficiently with excellent complexity/performance trade-offs \cite{gesbert2007shifting}. 
Besides, MU-MIMO communications also require the accurate channel state information (CSI) to completely eliminate the IUI, which is mainly achieved by the channel estimation \cite{rao2014distributed,zeng2024tutorial}. However, both the overheads of the channel estimation and computational complexity for the IUI suppression are significant, especially when the antenna size becomes large and massive user access is allowed.




Numerous efforts have been devoted to addressing the IUI issue for MU-MIMO systems since the advent of multiple-input multiple-output (MIMO) technology \cite{gesbert2007shifting}. For the early MIMO system, simple linear processing techniques only incorporate the zero forcing (ZF) and minimum mean squared error (MMSE) \cite{spencer2004introduction,spencer2004zero}. Otherwise, some complicated nonlinear approaches are required, such as the maximum-likelihood multi-user detection \cite{verdu1998multiuser} and ``dirty paper" coding \cite{costa1983writing,weingarten2006capacity}. As the MIMO technology gradually evolved towards the massive MIMO, typical antenna number at the base station (BS) increases to 64 or 128 \cite{bjornson2019massive,zhang2020prospective}. Benefited by such a boost in the antenna number, the most basic linear maximum-ratio combining (MRC) is able to achieve effective MU-MIMO communications with the suppressed IUI, thanks to approximately orthogonal channels between the BS and users \cite{marzetta2010noncooperative,ngo2013energy,rusek2012scaling}. Looking forward to the sixth-generation (6G) mobile communication networks, extremely large-scale multiple-input multiple-output (XL-MIMO) has been recognized as a promising technology to exceedingly improve the communication and sensing performance \cite{lu2024tutorial,lu2021communicating,bjornson2020power,hu2018beyond}. Specifically, super high beamforming gain and spatial resolution are perceived to be realized with the XL-MIMO, owing to its significant advance in the number of array elements, e.g., several hundreds or even thousands of antennas \cite{lu2024tutorial}. Moreover, larger array aperture also renders conventional uniform plane wave (UPW) model invalid, and more accurate uniform spherical wave (USW) propagation needs to be considered. Note that such a new characteristic is in fact beneficial to MU-MIMO communications since it can not only enable more users to access the BS, but also provide a new degree of freedom (DoF) to suppress the IUI by the distance separation \cite{lu2021near}. However, different from the multi-user massive MIMO, the performance of MRC in the XL-MIMO system may deteriorate because users are spatially correlated in both the angle and distance. Although ZF and MMSE techniques can exhibit superior effects, the sharp increase in both the array size and user number may cause super high complexity of calculating the matrix inversion \cite{wang2024tutorial}. 

There have been extensive works on investigating various linear methods to pursue the desirable complexity reduction. In particular, the beamspace processing has been regarded as an effective way in the array signal processing by transforming the original signal space into a subspace with lower dimension \cite{krim1996two,xu1994beamspace,zoltowski1993beamspace,zoltowski1996closed}. 
Besides, such a concept has been widely studied in the beamspace MIMO for multi-user communications, such as the channel modeling \cite{brady2013beamspace,sayeed2002deconstructing}, channel estimation \cite{guo2017millimeter,zhang2021gridless}, and beam selection \cite{amadori2015low,gao2016near}.
Furthermore, convolutional beamspace (CBS) has recently emerged as a novel beamspace technique \cite{chen2020convolutional,chen2021distributed,vaidyanathan2020convolutional}. Unlike the commonly used discrete Fourier transform (DFT)-based beamspace in the existing literature, the CBS is able to preserve the Vandermonde structure of array outputs, which is thus convenient for subsequent operations. Note that such a distinct property has been initially leveraged for both the localization and channel estimation. On the one hand, several subspace-based super-resolution algorithms, such as mutiple signal classification (MUSIC) \cite{schmidt1986multiple} and root-MUSIC \cite{rao1989performance} can be directly performed for the high-resolution direction of arrival (DoA) estimation \cite{chen2020convolutional}. On the other hand, the channel estimation can be achieved in a low-complexity manner, since such an issue can be transformed into a problem similar to the DoA estimation, especially in millimeter wave (mmWave) scenarios \cite{chen2024channel}. Another important property of the CBS transformation is that original array outputs will be equivalently convolved with taps of a spatial domain digital filter, e.g., finite impulse response (FIR) or infinite impulse response (IIR) \cite{chen2022convolutional}, thus yielding spatial filtering effects. Consequently, interference targets in the DoA estimation will be partially filtered, which is in general beneficial to both the estimation accuracy and computational complexity.
Inspired by this, the CBS can also be exploited as a preprocessing step to assist the IUI mitigation for MU-MIMO communications, which also owns a similar formulation with the DoA estimation. Specifically, interference signals from different users can be first weakened via the spatial filtering of the CBS, which can significantly simplify the subsequent IUI suppression. Moreover, given that the Vandermonde structure of array outputs is still maintained, it is easy to carry out classical linear processing schemes, i.e., MRC, ZF and MMSE, based on the dimension-reduced matrix.

It is worth mentioning that the existing literature on CBS mainly focuses on the conventional far-field UPW model.
Intuitively, the ability to alleviate the IUI via the CBS technique can be enhanced as the array aperture becomes larger. This is expected since the spatial filtering performance of the CBS significantly depends on the spatial domain digital filter's length, which has to be smaller than the element number of the array \cite{chen2020convolutional}. Specifically, longer filter's length is required to filter more interference signals or bring larger attenuation for interference targets. Hence, to achieve more desirable IUI mitigation, the array size is likely to increase by at least an order of magnitude, corresponding to the XL-MIMO, such that the near-field USW model cannot be neglected. In this case, the useful Vandermonde structure of array outputs may vanish by taking into account spherical wavefronts, thus rendering classical tools for designing the digital filter no longer valid. However, such an issue with the ``filter-like" structure can still be addressed by formulating an optimization problem, which aims at minimizing the IUI and obtaining an optimal CBS filter \cite{davidson2010enriching}.

Therefore, we study in this paper the low-complexity CBS-based beamforming scheme in MU-MIMO communications using the extremely large-scale array, where both the far-field UPW and near-field USW models are investigated. The main contributions of this paper are summarized as follows.
\begin{itemize}
    \item Firstly, classical linear beamforming schemes including MRC, ZF and MMSE are introduced based on the generic near-field modeling for MU-MIMO communications. To gain useful insights, the CBS preprocessing is first applied to MU-MIMO systems with the far-field UPW approximation. Furthermore, the CBS-based MRC beamforming is proposed, where trade-offs between its performance of the IUI mitigation and beamforming gain are revealed based on the equivalent beam pattern. Specifically, obtaining better IUI mitigating effects may sacrifice the beamforming gain of the whole array. Besides, the corresponding complexity analysis of the CBS-based MRC beamforming is provided by comparison with classical MRC, ZF and MMSE schemes. It is shown that the CBS-based MRC beamforming can outperform aforementioned linear schemes in terms of the performance or complexity.
    \item Next, the CBS-based MMSE beamforming is proposed, which aims at lowering the unfavorable complexity of the classical MMSE scheme by filtering the majority of interference users. It is shown that the CBS-based MMSE beamforming achieves a near-optimal performance with lower complexity. Furthermore, the array decimation is utilized to further reduce the computational complexity since only few users will be maintained with the aid of the CBS preprocessing. In particular, larger decimated length may result in smaller complexity by losing the performance of the IUI suppression. Besides, the aforementioned beamforming is mainly performed under the non-white noise brought by the CBS. Hence, the impact of the noise whitening is also investigated in both the additional complexity and beamforming design.
    \item Lastly, the CBS-based beamforming is extended to the near-field case, where users may be located in the near-field region of the array. To gain more insights into the beamforming design, the second-order Taylor approximation is considered when characterizing the phase term in the channel modeling. To resolve optimal near-field filtering coefficients, we first introduce a metric for each user, termed post-CBS vector, whose Euclidean norm accounts for its signal intensity. Then, an optimization problem is formulated based on the designed passband, stopband and transition band. Furthermore, such an issue can be transformed into a basic convex problem by using slack variables and successive convex approximation (SCA). Numerical results are presented to demonstrate the effectiveness of our proposed beamforming method.
\end{itemize}

The rest of this paper is organized as follows. Section \uppercase\expandafter{\romannumeral2} introduces the near-field modeling for MU-MIMO communications and presents three classical linear beamforming schemes. In Section \uppercase\expandafter{\romannumeral3}, the CBS preprocessing is presented based on the far-field model and the CBS-based MRC beamforming is then proposed. Furthermore, Section \uppercase\expandafter{\romannumeral4} analyzes the CBS-based MMSE beamforming scheme. We then extend the CBS-based beamforming into the near-field case in Section \uppercase\expandafter{\romannumeral5}.  Numerical results are provided in Section \uppercase\expandafter{\romannumeral6}. Finally, we conclude this paper in Section \uppercase\expandafter{\romannumeral7}.

\textit{Notations:} $\mathbb{C}^{M \times N}$ denotes the space of $M \times N$ complex-valued matrices. For a vector $\mathbf{z}$, $\|\mathbf{z}\|$ and $[\mathbf{z}]_i$ represent its Euclidean norm and $i$-th element, respectively. For a complex number $x$, $|x|$ denotes it absolute value. $\lceil \cdot \rceil$ refers to the ceil operation for the real number. $\setminus$ is the set minus operation. $\mathcal{CN} \left(\mathbf{0},\mathbf{\Sigma}\right)$ is the distribution of a circularly symmetric complex Gaussian (CSCG) random vector with mean $\mathbf{0}$ and covariance matrix $\mathbf{\Sigma}$. $\mathbb{E}\left[\cdot\right]$ denotes the statistical expectation.


\section{System Model}
\begin{figure}[!t]
	\centering
	\includegraphics[width=7.0cm]{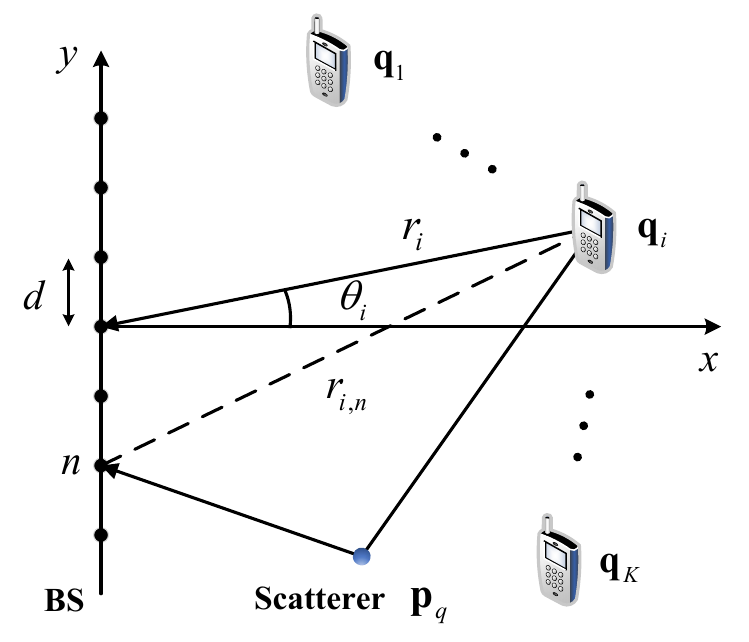}
	\caption{Multi-user communication with a ULA.}
	\label{system model}
\end{figure}

As shown in Fig. \ref{system model}, we consider an uplink communication system, where the BS serves $K$ single-antenna users. Moreover, the BS is equipped with a uniform linear array (ULA) with $N$ antenna elements. The separation between adjacent antenna elements is denoted by $d$. Without loss of generality, the ULA is placed along the $y$-axis and centered at the origin. For notational convenience, we assume that $N$ is an odd number. Therefore, the location of the $n$-th array element is denoted by $\mathbf{w}_n=\left[0,nd\right]^T$, where $n=0,\pm 1,\cdots,\pm \left(N-1\right)/2$. Denote the distance between the $i$-th user and the array center as $r_i$, $1 \leq i \leq K$, and the location of user $i$ is thus given by $\mathbf{q}_i=\left[r_i \cos\theta_i,r_i \sin\theta_i\right]^T$, where $\theta_i \in \left[-\frac{\pi}2,\frac{\pi}2\right]$ denotes the direction of $\mathbf{q}_i$ when viewed from the origin. Based on this, the distance between user $i$ and the $n$-th array element can be expressed as
\begin{equation}
  \label{distance_user_element}
    r_{i,n}=\left\|\mathbf{q}_i-\mathbf{w}_n\right\|
    =r_i\sqrt{1-2n\varepsilon_i \sin \theta_i +n^2\varepsilon_i^2},
\end{equation}
where $\varepsilon_i \triangleq \frac{d}{r_i}$. Note that in practice, we have $\varepsilon_i \ll 1$. The $n$-th entry of the $N$-element array response vector $\mathbf{a}_N \left(r_i,\theta_i\right)$ for user $i$ is modeled as \cite{lu2024tutorial}
\begin{equation}
	\label{array response vector for uplink communication}
	\left[\mathbf{a}_N \left(r_i,\theta_i\right) \right]_n
	= \frac{\sqrt{\beta_0}}{r_{i,n}} e^{-j\frac{2\pi}{\lambda}r_{i,n}},
\end{equation}
where $\beta_0$ denotes the channel gain at the reference distance of $1\ \text{m}$ and $\lambda$ is the signal wavelength. Furthermore, the line-of-sight (LoS) channel matrix considering all the K users can be expressed as
\begin{equation}
\label{Vandermonde matrix and user information}
    \mathbf{A}_{\text{LoS}}=\left[\mathbf{a}_N \left(r_1,\theta_1\right),\cdots,\mathbf{a}_N \left(r_K,\theta_K\right)\right].
\end{equation}

We also consider the multi-path propagation with $Q$ scatterers, and the location of the $q$-th scatterer is then denoted by $\mathbf{p}_{q}=[\tilde{r}_q\cos\tilde{\theta}_q,\tilde{r}_q\sin\tilde{\theta}_q]^T$. Note that $\tilde{r}_q$ and $\tilde{\theta}_q$ account for the distance between the $q$-th scatterer and the array center, and the direction of $\mathbf{p}_q$ with respect to the origin, respectively. Thus, the non-line-of-sight (NLoS) array response vector for user $i$ can be expressed as
\begin{equation}
    \label{NLoS channel vector for user i}
    \mathbf{a}_N^{\text{NLoS}} \left(r_i,\theta_i\right) = \sum_{q=1}^Q \beta_{q,i} \mathbf{b}_N \left(\tilde{r}_q,\tilde{\theta}_q\right) e^{-j\frac{2\pi}{\lambda} \left\|\mathbf{p}_{q}-\mathbf{q}_i\right\|},
\end{equation}
where $\beta_{q,i}$ denotes the complex-valued gain of the $q$-th NLoS link for user $i$, which includes both the path loss and random scattering
coefficient. Besides, the array response vector between the $q$-th scatterer and BS is 
\begin{equation}
    \label{NLoS array response vector for scatter q}
    \left[\mathbf{b}_N \left(\tilde{r}_q,\tilde{\theta}_q\right)\right]_n = \frac{\sqrt{\beta_0}}{\left\|\mathbf{p}_{q}-\mathbf{w}_n\right\|} e^{-j\frac{2\pi}{\lambda} \left\|\mathbf{p}_{q}-\mathbf{w}_n\right\|}.
\end{equation}
Similarly, the NLoS channel matrix is given by
\begin{equation}
\label{NLoS matrix}
    \mathbf{A}_{\text{NLoS}}=\left[\mathbf{a}_N^{\text{NLoS}} \left(r_1,\theta_1\right),\cdots,\mathbf{a}_N^{\text{NLoS}} \left(r_K,\theta_K\right)\right].
\end{equation}

Thus, the received signal at the BS can be expressed as 
\begin{equation}
    \label{received signal at the BS}
    \begin{aligned}
        \mathbf{y} 
        &= \mathbf{A} \mathbf{x} + \mathbf{n}\\
    &=\sum_{i=1}^K \left[\mathbf{a}_N \left(r_i,\theta_i\right)+\xi\mathbf{a}_N^{\text{NLoS}}\left(r_i,\theta_i\right)\right]\sqrt{P_i}x_i+\mathbf{n},
    \end{aligned}
\end{equation}
where $\mathbf{A} \triangleq \mathbf{A}_{\text{LoS}}+\xi\mathbf{A}_{\text{NLoS}}$ denotes the multi-path near-field channel matrix with $\xi \in \{0,1\}$ determining the existence of NLoS components, and
$\mathbf{x}=[\sqrt{P_1}x_1,\sqrt{P_2}x_2,$ $\cdots,\sqrt{P_K}x_K]^T$ is the transmit signal with $P_i$ and $x_i$ representing the transmit power and the information-bearing symbol of user $i$, respectively. Besides, $\mathbf{n} \in \mathbb{C}^{N \times 1} \sim \mathcal{CN}\left(0,\sigma^2\mathbf{I}_N\right)$ is the additive white Gaussian noise (AWGN) with $\sigma^2$ denoting its variance.

\begin{figure*}[!t]
	\centering
	\includegraphics[width=12.0cm]{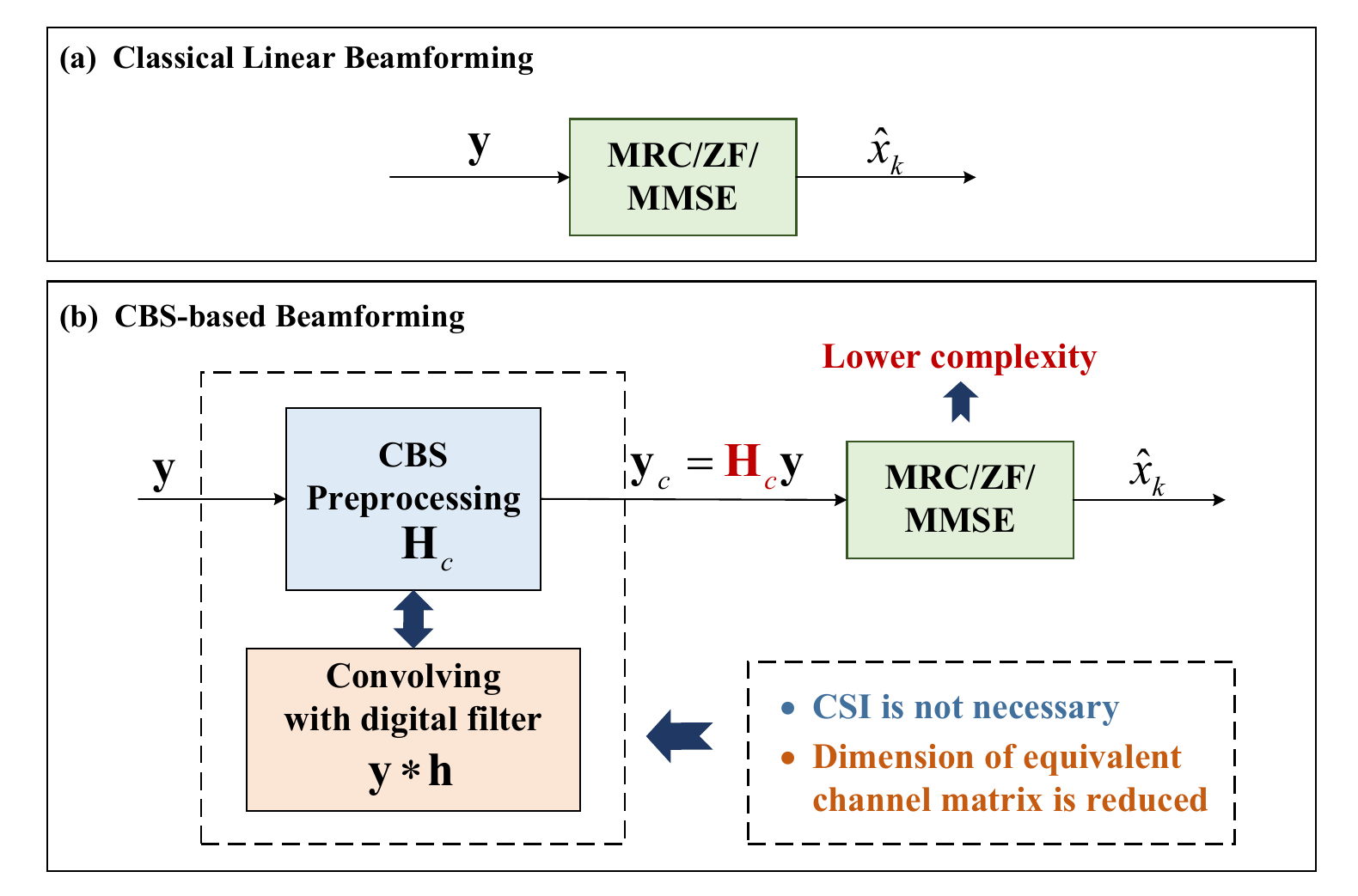}
	\caption{Illustrations for CBS-based beamforming versus classical linear beamforming schemes.}
	\label{CBS_vs_classical}
\end{figure*}

For convenience, we first consider the basic LoS-dominated channel, i.e., $\xi  =0$. In this case, to obtain the signal of user $k$, a linear receive beamforming vector  $\mathbf{v}_k \in \mathbb{C}^{N \times 1}$ is used, which satisfies $\left\|\mathbf{v}_k\right\|=1$. Therefore, the received signal at user $k$ can be further written as \cite{lu2021near}
\begin{equation} 
    \label{received signal after beamforming}
    \begin{aligned}
        y_k 
        &= \mathbf{v}_k^H \mathbf{a}_N \left(r_k,\theta_k\right)\sqrt{P_k} x_k \\
        & \quad+\sum_{i=1,i \neq k}^{K} \mathbf{v}_k^H \mathbf{a}_N \left(r_i,\theta_i\right)\sqrt{P_i} x_i 
        + \mathbf{v}_k^H \mathbf{n},
    \end{aligned}
\end{equation}
and its received signal-to-interference-plus-noise ratio (SINR) is given by
\begin{equation}
	\label{sinr of conventional beamforming}
	\gamma_k = \frac{\bar{P}_k\left|\mathbf{v}_k^H \mathbf{a}_N \left(r_k,\theta_k\right)\right|^2}{1+\sum_{i=1,i \neq k}^{K} \bar{P}_i \left| \mathbf{v}_k^H \mathbf{a}_N \left(r_i,\theta_i\right)\right|^2},
\end{equation}
where $\bar{P}_i \triangleq \frac{P_i}{\sigma^2}$ refers to the transmit signal-to-noise ratio (SNR) for user $i$. Thus, the achievable sum rate in bits/second/Hz (bps/Hz) is 
\begin{equation}
    \label{achievable sum rate}
    R_{\text{sum}} = \sum_{k=1}^K \log_2 \left(1+\gamma_k\right).
\end{equation}

Next, classical linear beamforming schemes for multi-user communications will be introduced based on \eqref{received signal after beamforming}.
Specifically, commonly used linear methods include MRC, ZF and MMSE beamforming, which are listed as follows \cite{ngo2013energy}:

\begin{itemize}
	\item \textbf{MRC beamforming:} The receive beamforming matrix is 
	\begin{equation}
		  \label{MRC beamforming matrix}
		  \mathbf{B}_{\text{MRC}} = \mathbf{A}^H.
        \end{equation}
        The linear beamforming vector for user $k$ can be obtained as
        \begin{equation}
            \label{MRC beamforming vector}
            \mathbf{v}_{\text{MRC},k} = \frac{\mathbf{a}_N \left(r_k,\theta_k\right)}{\left\|\mathbf{a}_N \left(r_k,\theta_k\right)\right\|}.
        \end{equation}
	Note that the beam pattern of $\mathbf{v}_{\text{MRC},k}$ with respect to an arbitrary observation location $\left(r,\theta\right)$ can be expressed as 
	\begin{equation}
		\label{beam pattern for MRC}
f_{\text{MRC}}^k\left(r,\theta\right)=\frac1{\sqrt{N}}\left|\mathbf{v}_{\text{MRC},k}^H \left(r_k,\theta_k\right) \mathbf{a}_N \left(r,\theta\right)\right|.
	\end{equation}
\end{itemize}


\begin{itemize}
	\item \textbf{MMSE beamforming:} The receive beamforming matrix is 
	\begin{equation}
		\label{MMSE beamforming matrix}
		\mathbf{B}_{\text{MMSE}} = \left(\mathbf{A}^H \mathbf{A}+\bar{\mathbf{P}}^{-1}\right)^{-1}\mathbf{A}^H.
	\end{equation}
        where $\bar{\mathbf{P}} \triangleq \text{diag}\left(\bar{P_1},\cdots,\bar{P}_K\right)$. Note that the \textbf{ZF beamforming} can be regarded as a special case of the MMSE by letting $\bar{P_1},\cdots,\bar{P}_K \rightarrow \infty$, thus omitted for brevity here. Besides, the linear MMSE beamforming vector for user $k$ can be obtained as \cite{lu2021near}
        \begin{equation}
            \label{MMSE beamforming vector}
            \mathbf{v}_{\text{MMSE},k} = \frac{\mathbf{C}_k^{-1}\mathbf{a}_N \left(r_k,\theta_k\right)}{\left\|\mathbf{C}_k^{-1}\mathbf{a}_N \left(r_k,\theta_k\right)\right\|},
        \end{equation}
        where $\mathbf{C}_k \triangleq \sum_{i=1,i \neq k}^K \bar{P}_i \mathbf{a}_N \left(r_i,\theta_i\right) \mathbf{a}_N \left(r_i,\theta_i\right)^H+\mathbf{I}_N$ denotes the interference-plus-noise covariance matrix.
\end{itemize}
It is worth mentioning that the MRC beamforming is easy to implement, while its performance of the IUI suppression significantly depends on the antenna number $N$. By contrast, ZF and MMSE beamforming schemes tend to exhibit superior outcomes by losing the receive SNR \cite{lu2021near}, as well as increasing the computational complexity, which is related to the user number $K$. 
Based on beamforming matrices given by \eqref{MRC beamforming matrix} and \eqref{MMSE beamforming matrix}, it can be observed that
the complexity of performing the MRC beamforming is $\mathcal{O}\left\{KN\right\}$, while the complexity of using the ZF or MMSE is $\mathcal{O}\left\{K^2 N+K^3+KN+K^2\right\}$. Thus, ZF and MMSE beamforming schemes may require a significant amount of computational resources when the user number $K$ becomes very large.

\section{CBS-based beamforming}

\begin{figure*}[!t]
\normalsize
\setcounter{MYtempeqncnt}{\value{equation}}
\setcounter{equation}{\value{MYtempeqncnt}}

  
\begin{equation}
	\label{Toeplitz matrix}
	\begin{aligned}
		\mathbf{H}_c=
	\left[\begin{array}{cccccc}
		h_c(L-1) & \ldots & h_c(0) & 0 & \cdots & 0 \\
		0 & h_c(L-1) & \cdots & h_c(0) & \cdots & 0 \\
		\vdots & \vdots & \ddots & \vdots & \ddots & \vdots \\
		0 & 0 & \cdots & h_c(L-1) & \cdots & h_c(0)
	\end{array}\right]_{(N- L + 1) \times N}.
	\end{aligned}
\end{equation}
  
\setcounter{equation}{\value{MYtempeqncnt}+1}
\hrulefill
\vspace*{4pt}
\end{figure*}

Unlike classical linear beamforming schemes, where the target user's signal is directly processed by applying the MRC/ZF/MMSE, we use the CBS technique to first preprocess the received signal $\mathbf{y}$ in \eqref{received signal at the BS}, then followed by classical linear beamforming schemes, as shown in Fig. \ref{CBS_vs_classical}. It will be subsequently shown that the dimension of equivalent channel matrix can be obviously reduced. Besides, the CSI is indeed not required for the CBS preprocessing since the matrix $\mathbf{H}_c$ in Fig. \ref{CBS_vs_classical} can be designed in advance. Hence, the cost of both the channel estimation and beamforming design is expected to be significantly lowered owing to the dimension reduction. To clearly illustrate such potentials, we first consider the far-field case, where the distance in \eqref{distance_user_element} can be approximated as $r_{i,n}\approx r_i\left(1-n\varepsilon_i \sin \theta_i\right)$ by utilizing the first-order Taylor expansion. As such, the array response vector for user $i$ in \eqref{array response vector for uplink communication} can be simplified as
\begin{equation}
    \label{far-field array response vector}
    \left[\mathbf{a}_N \left(r_i,\theta_i\right) \right]_n \approx
    \left[\mathbf{a}_N \left(\omega_i\right) \right]_n
	= \frac{\sqrt{\beta_0}}{r_i} e^{-j\frac{2\pi}{\lambda}r_i} e^{jn\omega_i},
\end{equation}
where $\omega_i \triangleq \frac{2\pi}{\lambda}d \sin\theta_i$ can be interpreted as the spatial frequency corresponding to the incident angle $\theta_i$. In this case, the LoS channel component given by \eqref{Vandermonde matrix and user information} becomes a common Vandermonde matrix. By further assuming the half-wavelength antenna spacing, i.e., $d=\frac{\lambda}2$, we can obtain $\omega_i = \pi \sin\theta_i$.

\subsection{CBS preprocessing}

The CBS preprocessing attempts to achieve the  suppression of undesired signals,
by convolving the received signal vector $\mathbf{y}$ with a spatial domain FIR filter $h_c \left(l\right)$, $0 \leq l \leq L-1$, with $L$ denoting the length of the filter \cite{chen2020convolutional}. In fact, the output of the CBS preprocessing can also be expressed as multiplying the received signal $\mathbf{y}$ by a $(N- L + 1) \times N$ banded Toeplitz matrix $\mathbf{H}_c$, given by \eqref{Toeplitz matrix} shown at the top of this page.
Thus, the output signal can be expressed as
\begin{equation}
	\label{steady state output after CBS}
	\begin{aligned}
		\mathbf{y}_c
		&=\mathbf{H}_c \mathbf{y}
		=\mathbf{H}_c \mathbf{A} \mathbf{x} + \mathbf{H}_c \mathbf{n}\\
        &=\sum_{i=1}^{K} \mathbf{H}_c 
		\mathbf{a}_N \left(\omega_i\right)
		\sqrt{P_i} x_i 
		+ \mathbf{H}_c \mathbf{n},
	\end{aligned}
\end{equation}
which can be further expanded as \eqref{expanded output signal after CBS}, shown at the top of the next page. Note that $H_c\left(e^{j\omega}\right)=\sum_{l=0}^{L-1} h_c \left(l\right) e^{-j \omega l}$ in \eqref{expanded output signal after CBS} denotes the spatial frequency response of the filter $h_c\left(l\right)$, and $\mathbf{a}_{N-L+1} \left(\omega_i\right)$ is the first $N-L+1$ elements of the array response vector $\mathbf{a}_N \left(\omega_i\right)$. In particular, if $\omega_i$ lies in the stopband of the filter $H_c\left(e^{j\omega}\right)$, we have $\left|H_c\left(e^{j\omega_i}\right) \right|\approx 0$.
\begin{figure*}[!t]
\normalsize
\setcounter{MYtempeqncnt}{\value{equation}}
\setcounter{equation}{\value{MYtempeqncnt}}

  
\begin{equation}
	\label{expanded output signal after CBS}
	\begin{aligned}
		\mathbf{y}_c
		=\underbrace{e^{j\left(L-1\right)\omega_k} H_c\left(e^{j\omega_k}\right)
		\mathbf{a}_{N-L+1} \left(\omega_k\right)\sqrt{P_k} x_k}_{\text{user $k$'s signal}}
		+\underbrace{\sum_{i=1,i \neq k}^{K} e^{j\left(L-1\right)\omega_i} H_c\left(e^{j\omega_i}\right) \mathbf{a}_{N-L+1} \left(\omega_i\right)\sqrt{P_i} x_i}_{\text{interference signals}}
		+ \mathbf{H}_c \mathbf{n}.
	\end{aligned}
\end{equation}
  
\setcounter{equation}{\value{MYtempeqncnt}+1}
\hrulefill
\vspace*{4pt}
\end{figure*}
According to \eqref{expanded output signal after CBS}, the resulting signal $\mathbf{y}_c$ incorporates the user $k$'s signal, interference signals and noise. Thus, for user $k$, the most ideal case is that all the interference signals can be filtered, i.e., $\left|H_c\left(e^{j\omega_i}\right)\right| \rightarrow 0$ for $i \neq k$, and its own signal is perfectly remained, i.e., $\left|H_c\left(e^{j\omega_k}\right)\right| \rightarrow 1$.

\subsection{MRC beamforming}
It can be observed that 
the MRC beamforming can still be used to obtain the target signal
in \eqref{expanded output signal after CBS}. Compared with the classical MRC \eqref{MRC beamforming matrix},
the only difference is that the corresponding beamforming matrix  needs to be modified as
\begin{equation}
	\label{MRC matrix for CBS processing}
	\mathbf{B}_{c,\text{MRC}} = \mathbf{U}^H,
\end{equation}
where $\mathbf{U} \triangleq \mathbf{H}_c \mathbf{A}$.

    Besides, the linear CBS-based MRC beamforming vector for user $k$ can be expressed as 
    \begin{equation}
        \label{MRC vector for CBS processing}
        \mathbf{v}_{c,\text{MRC},k}=\frac{e^{j\left(L-1\right)\omega_k} H_c\left(e^{j\omega_k}\right) \mathbf{a}_{N-L+1} \left(\omega_k\right)}{\left\|e^{j\left(L-1\right)\omega_k} H_c\left(e^{j\omega_k}\right)\mathbf{a}_{N-L+1} \left(\omega_k\right)\right\|}.
    \end{equation}



Note that for the CBS-based MRC beamforming, the most ideal filter $H_c\left(e^{j\omega}\right)$ can not only retain the signal from user $k$ without any power loss, but also reject the IUI efficiently. 
On the other hand, due to the transformation from $\mathbf{a}_N \left(\omega_k\right)$ to $\mathbf{a}_{N-L+1} \left(\omega_k\right)$, the maximum beamforming gain of the system will be reduced from $N$ to $\left(N-L+1\right)$. In particular, an ideal filter, owning the narrower transition band and higher stopband attenuation, requires very large $L$. For the extreme case with $L=N$, the maximum beamforming gain will reduce to $1$. Therefore, how to design a spatial FIR filter with appropriate length is a critical problem for the CBS-based beamforming.

The output of the CBS-based MRC beamforming can be written as 
\begin{equation}
	\label{received signal after CBS}
	\begin{aligned}
		y_{c,k} &= \mathbf{v}_{c,\text{MRC},k}^H \mathbf{H}_c 
		\mathbf{a}_N \left(\omega_k\right) \sqrt{P_k} x_k\\
		&+\sum_{i=1,i \neq k}^{K} \mathbf{v}_{c,\text{MRC},k}^H \mathbf{H}_c 
		\mathbf{a}_N \left(\omega_i\right) \sqrt{P_i} x_i
		+\mathbf{v}_{c,\text{MRC},k}^H \mathbf{H}_c \mathbf{n}.
	\end{aligned}
\end{equation}
where the power of the resulting noise $n_c\triangleq\mathbf{v}_{c,\text{MRC},k}^H \mathbf{H}_c \mathbf{n}$ is given by \footnote{The noise whitening is not taken into account here, which will be subsequently clarified in detail}
\begin{equation}
	\label{autocorrelation of the noise after CBS}
	R_{n_c} = \mathbb{E}\left[n_c n_c^H\right]
	=\sigma^2 \mathbf{v}_{c,\text{MRC},k}^H \mathbf{H}_c \mathbf{H}_c^H \mathbf{v}_{c,\text{MRC},k}.
\end{equation}

\begin{figure*}[!t]
\normalsize
\setcounter{MYtempeqncnt}{\value{equation}}
\setcounter{equation}{\value{MYtempeqncnt}}

  
\begin{equation}
	\label{sinr after CBS}
	\begin{aligned}
	\gamma_{c,k}
    &=\frac{\bar{P}_k\left|\mathbf{v}_{c,\text{MRC},k}^H \mathbf{H}_c 
		\mathbf{a}_N \left(\omega_k\right)\right|^2}{\mathbf{v}_{c,\text{MRC},k}^H \mathbf{H}_c \mathbf{H}_c^H \mathbf{v}_{c,\text{MRC},k}+\sum_{i=1,i \neq k}^K \bar{P}_i \left|\mathbf{v}_{c,\text{MRC},k}^H \mathbf{H}_c 
		\mathbf{a}_N \left(\omega_i\right)\right|^2}\\
        &=\frac{\bar{P}_k\left|\mathbf{v}_{c,\text{MRC},k}^H 
		\mathbf{a}_{N-L+1} \left(\omega_k\right)\right|^2}{\mathbf{v}_{c,\text{MRC},k}^H \mathbf{H}_c \mathbf{H}_c^H \mathbf{v}_{c,\text{MRC},k}/\left|H_c\left(e^{j\omega_k}\right)\right|^2+\sum_{i=1,i \neq k}^{K}\bar{P}_i \left| \mathbf{v}_{c,\text{MRC},k}^H 
		\mathbf{a}_{N-L+1} \left(\omega_i\right)\frac{H_c\left(e^{j\omega_i}\right)}{H_c\left(e^{j\omega_k}\right)}\right|^2}.
	\end{aligned}
\end{equation}
  
\setcounter{equation}{\value{MYtempeqncnt}+1}
\hrulefill
\vspace*{4pt}
\end{figure*}

\begin{figure*}[!t]
  \centering
  \subfloat[$L=30$]{\includegraphics[width=8cm]{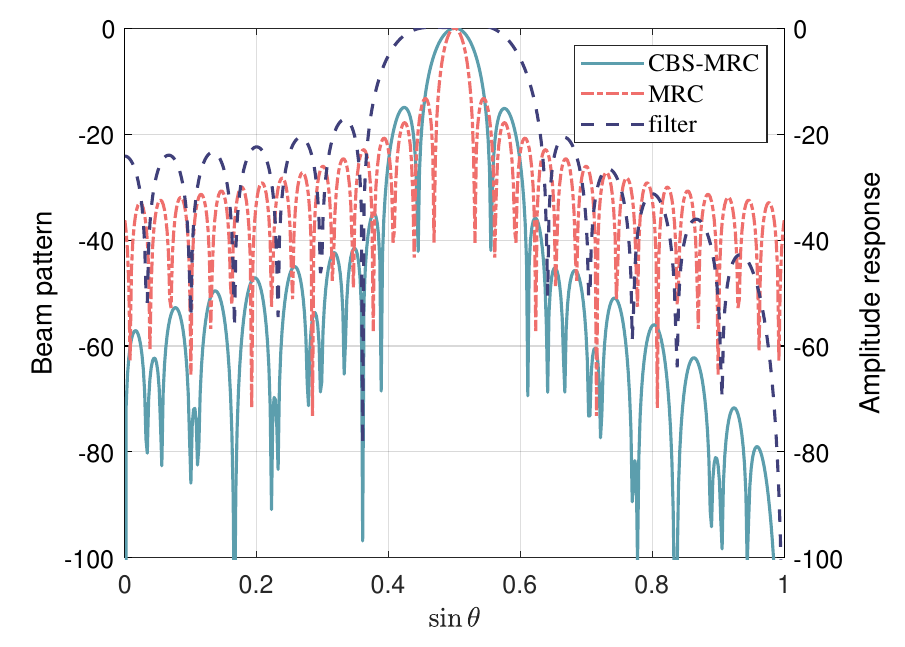}
  \label{bandpass_filter_information_L_30}}
  \hfil
  \subfloat[$L=64$]{\includegraphics[width=8cm]{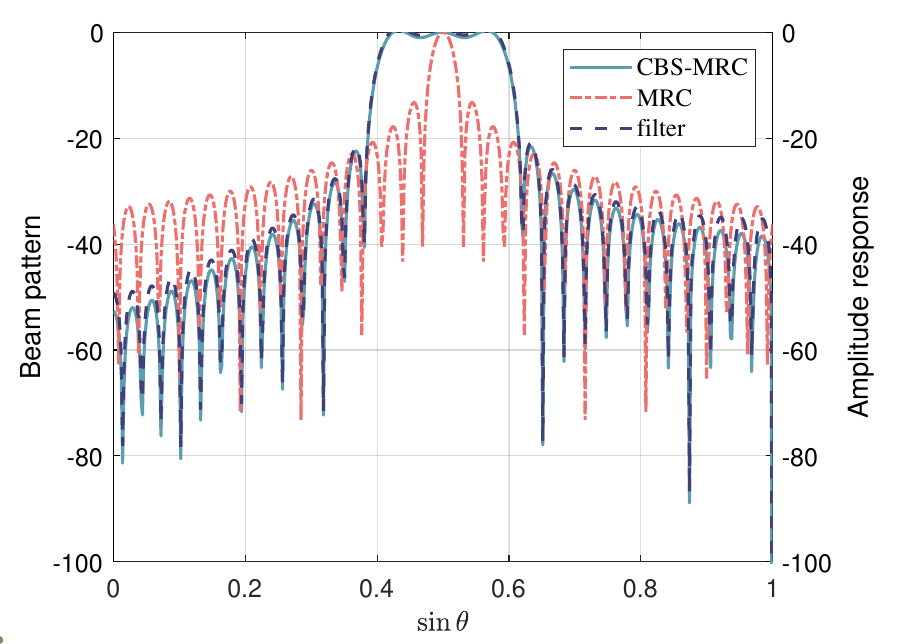}
  \label{bandpass_filter_information_L_64}}
  \caption{Beam patterns of CBS-based MRC versus conventional MRC, where we use $\sin\theta=\omega/\pi$ to characterize the spatial angular frequency $\omega$.}
  \label{Beam patterns of CBS-MRC and MRC}
\end{figure*}

Furthermore, the SINR of the CBS-based MRC beamforming can be expressed as \eqref{sinr after CBS}, shown at the top of the next page. 
Based on \eqref{sinr after CBS}, the equivalent beam pattern of $\mathbf{v}_{c,\text{MRC},k}$ with an arbitrary observation frequency $\omega$ can be obtained as 
\begin{equation}
	\label{beam pattern for CBS beamforming}
        \begin{aligned}
	f_{\text{CBS}}^k\left(\omega\right)
        &=\frac{\left|\mathbf{v}_{c,\text{MRC},k}^H 
		\mathbf{a}_{N-L+1} \left(\omega\right)\frac{H_c\left(e^{j\omega}\right)}{H_c\left(e^{j\omega_k}\right)}\right|}{\left\|\mathbf{v}_{c,\text{MRC},k}\right\| \left\|\mathbf{a}_{N-L+1} \left(\omega\right)\right\|}\\
        &=\underbrace{\frac{\left|\mathbf{v}_{c,\text{MRC},k}^H  \mathbf{a}_{N-L+1} \left(\omega\right)\right|}{\sqrt{N-L+1}} }_{\text{MRC beam pattern}} \cdot
        \underbrace{\left|\frac{H_c\left(e^{j\omega}\right)}{H_c\left(e^{j\omega_k}\right)}\right|}_{\text{filtering effect}}.
        \end{aligned}
\end{equation} 
For the conventional MRC beamforming \eqref{beam pattern for MRC}, its beam pattern under the far-field UPW  model is
\begin{equation}
    \label{beam pattern for MRC far-field}
f_{\text{MRC}}^k\left(\theta\right)=\frac1{\sqrt{N}}\left|\mathbf{v}_{\text{MRC},k}^H \left(\theta_k\right) \mathbf{a}_N \left(\theta\right)\right|.
\end{equation}
Therefore, it can be observed that the equivalent beam pattern of $\mathbf{v}_{c,\text{MRC},k}$ \eqref{beam pattern for CBS beamforming}  incorporates the beam pattern of the conventional MRC with $N-L+1$ array elements in \eqref{beam pattern for MRC far-field}, as well as the spatial amplitude-frequency response of the FIR filter.

Fig. \ref{Beam patterns of CBS-MRC and MRC} plots the equivalent beam patterns of the CBS-based MRC beamforming, where the filter length is set as $L=30$ and $L=64$, respectively. The beam patterns of conventional MRC beamforming are also shown based on the identical array with $N=65$. Note that the passband of the spatial FIR filter is designed as $\omega \in \left[0.4\pi,0.6\pi\right]$. Compared with the conventional MRC shown in Fig. \ref{Beam patterns of CBS-MRC and MRC}\subref{bandpass_filter_information_L_30}, the side lobes of the equivalent beam pattern with respect to the CBS-based MRC beamforming can be significantly suppressed. As a trade-off, the main lobe of the CBS-based MRC beamforming becomes wide. Furthermore,
it can be observed that longer filter length may cause wider main lobe.
In particular, when $L\rightarrow N$, the equivalent beam pattern of the CBS-based MRC beamforming becomes consistent with the amplitude-frequency response of the spatial filter, as shown in Fig. \ref{Beam patterns of CBS-MRC and MRC}\subref{bandpass_filter_information_L_64}.
Therefore, severe IUI may occur when many users are located in the passband of the filter. This validates the importance of designing an appropriate filter for the CBS-based beamforming, i.e., the trade-off between the filtering performance and beamforming gain.

In practice, there have been numerious methods to design a digital FIR filter, such as the \emph{Parks-McClellan Algorithm} \cite{mitra2006digital}. To illustrate clearly, we only use the basic \emph{Window Method} to generate a casual bandpass FIR filter based on its performance requirements. Then, the filter coefficients are given by
\begin{equation}
	\label{FIR bandpass filter casual form}
	h\left(l\right)
	=\frac{\sin\left[\omega_{c2} \left(l-M_L\right)\right]}{\pi \left(l-M_L\right)}-\frac{\sin\left[\omega_{c1} \left(l-M_L\right)\right]}{\pi \left(l-M_L\right)},
\end{equation}
where $L = 2M_L+1$. Besides, $\omega_{c1}$ and $\omega_{c2}$ denote the low and high cut-off frequencies, respectively. Note that the filtering performance of \eqref{FIR bandpass filter casual form} only depends on cut-off frequencies and the filter's length, which can be designed for users in advance. This renders that the complexity of the filter design does not need to be considered in the CBS-based beamforming.

The complexity of different linear beamforming schemes mentioned above are listed in TABLE \ref{Complexity for MRC, ZF, MMSE and CBS}, which mainly focuses on the number of multiplications during the signal processing. In particular, the computational complexity of the CBS-based MRC beamforming includes two parts, i.e., obtaining the resulting signal $\mathbf{y}_c=\mathbf{H}_c \mathbf{y}$ and performing the MRC beamforming for all the $K$ users. Note that numerical values of the above complexity analysis are provided in Section \uppercase\expandafter{\romannumeral6}. It can be observed that the complexity of the CBS-based MRC beamforming is similar to the conventional MRC by choosing suitable values for the filter's length $L$, but it achieves superior ability to suppress the IUI due to the spatial FIR filter. On the other hand, compared with the ZF and MMSE, the CBS-based MRC beamforming can significantly reduce the computational complexity, especially when the user number $K$ becomes very large.

\begin{table*}[!t]
	\centering
	\caption{Complexity of MRC, ZF, MMSE and CBS-based MRC.} 
	\label{Complexity for MRC, ZF, MMSE and CBS}
	\begin{tabular}{@{}lll@{}}
	\toprule
	Beamforming Type & Resulting Signal                                                                          & Complexity                                                    \\ \midrule
	MRC         & $\mathbf{A}^H \mathbf{y}$                                        & $\mathcal{O}\left\{KN\right\}$                                  \\ 
	ZF          & $\left(\mathbf{A}^H \mathbf{A}\right)^{-1} \mathbf{A}^H \mathbf{y}$                     & $\mathcal{O}\left\{K^2 N+K^3+KN+K^2\right\}$                    \\ 
	MMSE        & $\left(\mathbf{A}^H \mathbf{A}+\bar{\mathbf{P}}^{-1}\right)^{-1} \mathbf{A}^H \mathbf{y}$ & $\mathcal{O}\left\{K^2 N+K^3+KN+K^2\right\}$                    \\ 
	CBS-MRC     & $\mathbf{U}^H \mathbf{H}_c \mathbf{y}$                     & $\mathcal{O}\left\{\left(K+L\right)\left(N-L+1\right)\right\}$ \\  \bottomrule
	\end{tabular}
\end{table*}

\section{CBS-based MMSE beamforming}

In uplink multi-user communications, the MMSE beamforming is an optimal linear solution to maximize the SINR. However, the computational complexity of the MMSE beamforming is proportional to $K^3$, as shown in TABLE \ref{Complexity for MRC, ZF, MMSE and CBS}. That is to say, as the user number significantly increases, the corresponding complexity will become extremely high. Fortunately, the CBS processing can provide a new opportunity to address this issue, since users out of the passband are negligible after passing through the spatial FIR filter. In this case, the actual number of users that require the MMSE beamforming can be significantly reduced.

\subsection{Beamforming design}

Assume that the spatial FIR filter $H_c\left(e^{j\omega}\right)$ in \eqref{steady state output after CBS} is perfect 
and only $K_p$ users are located within the passband, whose angle set is denoted by $\Omega_p$. Let $\Lambda_p \triangleq \{i \big | \omega_i \in \Omega_p\}$ denote the index set of all passband users. Therefore, the corresponding channel matrix can be obtained as
\begin{equation}
    \label{channle matrix of users lying in the passband}
    \begin{aligned}
    \mathbf{A}_p = \mathbf{A} \mathbf{\Phi}_p,
    \end{aligned}
\end{equation}
where $\mathbf{\Phi}_p \in {\{ 0,1\} ^{K \times {K_p}}}$ is a transformation matrix with 
“1” only appearing once in each column and in the $i$-th row, $\forall i\in \Lambda_p $. Besides, the transmit signal with respect to all passband users can be expressed as $\mathbf{x}_p \triangleq \mathbf{\Phi}_p^H \mathbf{x}$.

\begin{figure*}[!t]
\normalsize
\setcounter{MYtempeqncnt}{\value{equation}}
\setcounter{equation}{\value{MYtempeqncnt}}

  
\begin{equation}
	\label{equivalent receive vector}
	\begin{aligned}
		\mathbf{y}_c 
        &\approx \mathbf{H}_c \mathbf{A}_p \mathbf{x}_p + \mathbf{H}_c \mathbf{n}
		=\sum_{i\in \Lambda_p} \mathbf{H}_c \mathbf{a}_N \left(\omega_i\right) \sqrt{P_i} x_i+\mathbf{H}_c \mathbf{n}\\
		&=\underbrace{e^{j\left(L-1\right)\omega_k} H_c\left(e^{j\omega_k}\right)
		\mathbf{a}_{N-L+1} \left(\omega_k\right)\sqrt{P_k} x_k}_{\text{user $k$'s signal}}
		+\underbrace{\sum_{i\in \Lambda_p,i \neq k} e^{j\left(L-1\right)\omega_i} H_c\left(e^{j\omega_i}\right) \mathbf{a}_{N-L+1} \left(\omega_i\right)\sqrt{P_i} x_i}_{\text{interference signals}}
		+ \mathbf{H}_c \mathbf{n}.
	\end{aligned}
\end{equation}
  
\setcounter{equation}{\value{MYtempeqncnt}+1}
\hrulefill
\vspace*{4pt}
\end{figure*}
As such, the resulting signal after the CBS preprocessing approximately equals to \eqref{equivalent receive vector}, shown at the top of the next page. Similarly, we first neglect the impact of the non-white noise during the beamforming design. Accordingly, the MMSE beamforming matrix given by \eqref{MMSE beamforming matrix} can be modified as
\begin{equation}
		\label{modefied MMSE beamforming matrix}
		\mathbf{B}_{c,\text{MMSE}} = \left(\mathbf{U}_p^H \mathbf{U}_p+\bar{\mathbf{P}}_p^{-1}\right)^{-1}\mathbf{U}_p^H,
\end{equation}
where $\mathbf{U}_p \triangleq \mathbf{H}_c \mathbf{A}_p$ and $\bar{\mathbf{P}}_p \triangleq \mathbf{\Phi}_p^H \bar{\mathbf{P}} \mathbf{\Phi}_p$.


    Besides, the linear CBS-based MMSE beamforming vector for user $k$ can be expressed as 
    \begin{equation}
	\label{CBS MMSE beamforming vector}
		\mathbf{v}_{c,\text{MMSE},k}
		= \frac{e^{j\left(L-1\right)\omega_k}\mathbf{C}_{c,k}^{-1} H_c\left(e^{j\omega_k}\right) \mathbf{a}_{N-L+1} \left(\omega_k\right)}{\left\|e^{j\left(L-1\right)\omega_k}\mathbf{C}_{c,k}^{-1} H_c\left(e^{j\omega_k}\right) \mathbf{a}_{N-L+1} \left(\omega_k\right)\right\|},
    \end{equation}
    where 
    \begin{align}
        &\mathbf{C}_{c,k}^{-1}
	=\mathbf{I}_{N-L+1} - \bar{\mathbf{U}}_{p,k} \left(\bar{\mathbf{P}}_{p,k}^{-1} +             \bar{\mathbf{U}}_{p,k}^H \bar{\mathbf{U}}_{p,k}\right)^{-1}\bar{\mathbf{U}}_{p,k}^H,\label{interference-plus-noise covariance matrix}\\
        &\bar{\mathbf{U}}_{p,k}
            \triangleq \mathbf{A}_L \mathbf{\Phi}_{p,k},\\
        &\bar{\mathbf{P}}_{p,k} 
        \triangleq \mathbf{\Phi}_{p,k}^H \bar{\mathbf{P}} \mathbf{\Phi}_{p,k}.
    \end{align}
    Note that by using the Woodbury matrix identity \cite{hager1989updating}, \eqref{interference-plus-noise covariance matrix} can be easily obtained based on the interference-plus-noise covariance matrix in \eqref{MMSE beamforming vector}. Besides,
    $\mathbf{A}_L$ denotes the first $N-L+1$ rows of the matrix $\mathbf{A}$, and we obtain $\mathbf{\Phi}_{p,k} \in {\{ 0,1\} ^{K \times (K_p-1)}}$ with “1” only appearing once in each column and in the $i$-th row, $i\in \Lambda_p \setminus k $. According to \eqref{CBS MMSE beamforming vector}, it can be observed that both the user number and equivalent array dimension are reduced, compared with the classical MMSE beamforming.


\subsection{Array decimation}

Besides the user number, the computational complexity of the classical MMSE beamforming also depends on the array dimension, which provides DoFs for the IUI mitigation. In practice, it is unnecessary to use all antenna elements to perform the MMSE \cite{veetil2015performance}, especially when the user number is small. Since only $K_p$ users remain after the CBS preprocessing, the complexity can be further reduced by performing the MMSE with a smaller array.
Based on this, the resulting vector \eqref{equivalent receive vector} can be uniformly decimated with the interval $N_d$. Hence, the decimation matrix can be obtained as 
\begin{equation}
	\label{decimation matrix}
	\mathbf{D}_\mu=\left[
	\boldsymbol{\delta}_\mu, \boldsymbol{\delta}_{\mu+N_d}, \cdots, \boldsymbol{\delta}_{\mu+(J-1) N_d}
	\right]^T,  0 \leq \mu \leq N_d-1,
\end{equation}
where $\boldsymbol{\delta}_i$ denotes the $i$-th standard orthogonal basis of the $\left(N-L+1\right)$ dimension space, $\mu$ is the starting point of the decimation, and $J \triangleq \lceil \frac{N-L+1}{N_d} \rceil$ accounts for the decimated array length. Therefore, the decimated resulting vector can be further expressed as \eqref{decimated receive vector}, shown at the top of the next page.

\begin{figure*}[!t]
	\normalsize
	\setcounter{MYtempeqncnt}{\value{equation}}
	\setcounter{equation}{\value{MYtempeqncnt}}
	
	  
	\begin{equation}
		\label{decimated receive vector}
	\begin{aligned}
		\mathbf{y}_{c,\mu}
		=\mathbf{D}_{\mu} \mathbf{y}_c
		\approx 
		\mathbf{D}_{\mu}\mathbf{H}_c \mathbf{a}_N \left(\omega_k\right) \sqrt{P_k}x_k
		+\sum_{i\in \Lambda_p,i \neq k} \mathbf{D}_{\mu}\mathbf{H}_c \mathbf{a}_N \left(\omega_i\right) \sqrt{P_i} x_i
		 +\mathbf{D}_{\mu}\mathbf{H}_c \mathbf{n}.
	\end{aligned}
	\end{equation}
	  
	\setcounter{equation}{\value{MYtempeqncnt}+1}
	\hrulefill
	\vspace*{4pt}
\end{figure*}

    Furthermore, the linear CBS-based MMSE beamforming vector by applying the uniform decimation for user $k$ in \eqref{CBS MMSE beamforming vector} can be expressed as 
    \begin{equation}
	\label{CBS MMSE beamforming vector after decimation}
	\mathbf{v}_{cd,\text{MMSE},k}
		= \frac{e^{j\left(L-1\right)\omega_k}\mathbf{C}_{cd,k}^{-1} \mathbf{D}_\mu H_c\left(e^{j\omega_k}\right) \mathbf{a}_{N-L+1} \left(\omega_k\right)}{\left\|e^{j\left(L-1\right)\omega_k}\mathbf{C}_{cd,k}^{-1} \mathbf{D}_\mu H_c\left(e^{j\omega_k}\right) \mathbf{a}_{N-L+1} \left(\omega_k\right)\right\|},
    \end{equation}
    where
\begin{equation}
	\label{CBS MMSE inverse after decimation}
	\mathbf{C}_{cd,k}^{-1}
	=\mathbf{I}_J - \bar{\mathbf{U}}_{p,d,k} \left(\bar{\mathbf{P}}_{p,k}^{-1} +             \bar{\mathbf{U}}_{p,d,k}^H \bar{\mathbf{U}}_{p,d,k}\right)^{-1}\bar{\mathbf{U}}_{p,d,k}^H,
\end{equation}
with $\bar{\mathbf{U}}_{p,d,k} \triangleq \mathbf{D}_\mu \bar{\mathbf{U}}_{p,k}$.

\begin{figure*}[!t]
	\normalsize
	\setcounter{MYtempeqncnt}{\value{equation}}
	\setcounter{equation}{\value{MYtempeqncnt}}
	
	  
	\begin{equation}
		\label{complexity for the CBS-MMSE beamforming}
		\begin{aligned}
			\mathcal{O}\left(L\left(N-L+1\right)+\sum_{m=1}^M\left[K_m^2\left(N-L+1\right)+K_m^3+K_m\left(N-L+1\right)+K_m^2\right]\right).
		\end{aligned}
	\end{equation}
	  
	\setcounter{equation}{\value{MYtempeqncnt}+1}
	\hrulefill
	\vspace*{4pt}
\end{figure*}

\subsection{Complexity analysis}

Note that the performance of the CBS-based MMSE beamforming is closely related to the spatial FIR filter design. Given the symmetry of the FIR filter, we only need to consider the positive half space of the spatial frequency, i.e., $\omega \in \left[0,\pi\right]$. To obtain an optimal FIR filter for a specific user, its passband cannot be too wide, thus rendering more interference signals filtered. As such, a filter bank composed of several filters with the identical passband width is needed to cover all users in the whole space. Thus, the spatial frequency domain will be divided into $M$ segments, and the CBS-based MMSE beamforming will be performed in each segment. For ease of exposition, let us denote the user number with respect to the $m$-th segment by $K_m$. The complexity of the CBS-based MMSE beamforming without any decimation is given by \eqref{complexity for the CBS-MMSE beamforming},
shown at the top of the next page. Specifically, both the complexity of obtaining the resulting signal $\mathbf{y}_c=\mathbf{H}_c \mathbf{y}$ and multiplications involved in the MMSE beamforming in each segment is taken into account.

To gain more useful insights, we assume that all users are uniformly distributed, i.e., the user number in each segment approximately equals to $K/M$. Based on this, both the complexity of the CBS-based MMSE beamforming with and without the uniform decimation is listed in TABLE \ref{Complexity for CBS-MMSE beamforming}. The corresponding numerical values are given in Section \uppercase\expandafter{\romannumeral6}.
For communication scenarios with massive user access, i.e., $K$ is very large, it can be shown that the CBS-based MMSE beamforming can obviously reduce the computational complexity, compared with the conventional ZF and MMSE. Specifically, the complexity of the ZF or MMSE beamforming is related to $K^3$. However, for the CBS-based MMSE beamforming, we only need to perform the MMSE on $K/M$ users every time by using the CBS technique, instead of $K$ in classical cases. 

\begin{table*}[!t]
	\centering
	\caption{Complexity of CBS-based MMSE beamforming.} 
	\label{Complexity for CBS-MMSE beamforming}
	\begin{tabular}{@{}lll@{}}
	\toprule
	Beamforming Type              & Complexity                                                                                                                                                                                                 \\ \midrule
	CBS-MMSE (no decimation) & $\mathcal{O}\left\{L\left(N-L+1\right)+M\left[\left(\frac{K}{M}\right)^2\left(N-L+1\right)+\left(\frac{K}{M}\right)^3+\left(\frac{K}{M}\right)\left(N-L+1\right)+\left(\frac{K}{M}\right)^2\right]\right\}$ \\ [2ex] 
	CBS-MMSE (decimation)    & $\mathcal{O}\left\{L\left(N-L+1\right)+M\left[\left(\frac{K}{M}\right)^2\frac{N-L+1}{N_d}+\left(\frac{K}{M}\right)^3+\left(\frac{K}{M}\right)\frac{N-L+1}{N_d}+\left(\frac{K}{M}\right)^2\right]\right\}$ \\ \bottomrule
	\end{tabular}
\end{table*}

\subsection{Impact of noise whitening}

As mentioned earlier, the MMSE beamforming is optimal only when the white noise exists. By introducing the Toeplitz matrix $\mathbf{H}_c$, the white Gaussian noise $\mathbf{n}$ becomes $\mathbf{H}_c \mathbf{n}$ in \eqref{steady state output after CBS}, which is obviously no longer white. To remove the impact of such non-white noise, we can first perform the noise whitening on the resulting vector $\mathbf{y}_c$ after the CBS preprocessing. By utilizing the whitening operator $\mathbf{W}_c=\left(\mathbf{H}_c \mathbf{H}_c^H\right)^{-1/2} \in \mathcal{C}^{\left(N-L+1\right)\times\left(N-L+1\right)} $ \cite{proakis2008digital}, the output vector after the noise whitening is represented as
\begin{equation}
	\label{outputs after the whitening filter}
	\begin{aligned}
		\mathbf{y}_w 
		&=\mathbf{W}_c \mathbf{y}_c =\mathbf{W}_c \mathbf{H}_c \mathbf{A} \mathbf{x}
		+\mathbf{W}_c \mathbf{H}_c \mathbf{n}\\
		&=\sum_{i=1}^{K} \mathbf{W}_c \mathbf{H}_c 
		\mathbf{a}_N \left(\omega_i\right)
		\sqrt{P_i}x_i + \mathbf{W}_c \mathbf{H}_c \mathbf{n}\\
		&=\sum_{i=1}^{K}  e^{j\left(L-1\right)\omega_i} H_c\left(e^{j\omega_i}\right) \mathbf{W}_c
		\mathbf{a}_{N-L+1} \left(\omega_i\right)\sqrt{P_i} x_i+\mathbf{n}_w,
	\end{aligned}
\end{equation}
where $\mathbf{n}_w \triangleq \mathbf{W}_c \mathbf{H}_c \mathbf{n} \in \mathcal{CN}\left(0,\sigma^2\mathbf{I}_{N-L+1}\right)$ denotes the equivalent Gaussian white noise. For convenience, let us define $\mathbf{G}_c \triangleq \mathbf{H}_c \mathbf{H}_c^H$, which is still a Toeplitz matrix with the first row $\left[g\left(0\right),g^{*}\left(1\right),g^{*}\left(2\right),\cdots,g^{*}\left(N-L\right)\right]$ \cite{chen2020convolutional}. Moreover, its each entry can be expressed as a convolution form, i.e., 
\begin{equation}
	\label{convolution form of the entry of first row}
	g\left(j\right)=\sum_l h\left(l\right) h^{*}\left(l-j\right),0\leq j \leq N-L.
\end{equation}
Note that the Toeplitz matrix $\mathbf{G}_c$ can be generated by simply shifting its first row. Therefore, the computational complexity of obtaining the matrix $\mathbf{G}_c$ is mainly determined by the number of multiplications between non-zero elements in computing $g\left(j\right)$, 
and thus given by $\mathcal{O}\left(L\left(L+1\right)/2\right)$.
Next, specific CBS-based beamforming schemes using the noise whitening will be provided based on \eqref{outputs after the whitening filter}.

\begin{itemize}
    \item For the CBS-based MRC beamforming with the noise whitening, the linear beamforming vector for user $k$ is given by
    \begin{equation}
		\label{CBS MRC beamforming vector white noise}
		\mathbf{v}_{cw,\text{MRC},k}
		=\frac{\mathbf{W}_c^{-1} e^{j\left(L-1\right)\omega_k} H_c\left(e^{j\omega_k}\right) \mathbf{a}_{N-L+1} \left(\omega_k\right)}{\left\|\mathbf{W}_c^{-1} e^{j\left(L-1\right)\omega_k} H_c\left(e^{j\omega_k}\right) \mathbf{a}_{N-L+1} \left(\omega_k\right)\right\|}.
    \end{equation}
\end{itemize}
	


\begin{itemize}
    \item For the CBS-based MMSE beamforming with the noise whitening, the beamforming matrix is given by
    \begin{equation}
		\label{CBS MMSE beamforming matrix white noise with users in bandpass}
		\mathbf{B}_{c,\text{MMSE},w}
		=\left(\mathbf{U}_{p,w}^H \mathbf{U}_{p,w}+\bar{\mathbf{P}}_p^{-1} \right)^{-1} \mathbf{U}_{p,w}^H ,
	\end{equation}
	where $\mathbf{U}_{p,w} \triangleq \mathbf{W}_c \mathbf{H}_c \mathbf{A}_p \in \mathbb{C}^{\left(N-L+1\right) \times K_p}$. Thus, the linear beamforming vector for user $k$ can be obtained as 
    \begin{equation}
        \label{cbs_mmse_vector_whitening_noise}
        \mathbf{v}_{cw,\text{MMSE},k} = \frac{\mathbf{b}_{c,\text{MMSE},w,k}}{\left\|\mathbf{b}_{c,\text{MMSE},w,k}\right\|},
    \end{equation}
    where $\mathbf{b}_{c,\text{MMSE},w,k}$ is the $k$-th column of $\mathbf{B}_{c,\text{MMSE},w}^H$.
\end{itemize}
	


Note that since the whitening operator $\mathbf{W}_c$ is introduced, it is challenging to obtain a similar form of the resulting vector as \eqref{equivalent receive vector}. In this case, the closed-form expression for the CBS-based MMSE beamforming vector \eqref{cbs_mmse_vector_whitening_noise} is not derived here. Besides, due to the special structure of the Toeplitz matrix $\mathbf{W}_c$, the complexity of computing its matrix inversion can be reduced to $\left(N-L+1\right)^2$, instead of $\left(N-L+1\right)^3$ \cite{strang1986proposal}. However, calculating the matrix $\mathbf{W}_c$ still requires at least $L\left(L+1\right)/2+\left(N-L+1\right)^2$ multiplications, which yields that the noise whitening may cause huge additional computational complexity. 


\section{Extension to the near-field case}

The CBS-based beamforming technique essentially uses $L-1$ DoFs for filtering the partial IUI and the remaining $N-L+1$ DoFs are used for the beamforming design. Since $L \leq N$, both the spatial filter and beamforming designs depend on the array size $N$. Specifically, larger array size may result in better performance of the IUI suppression by using the CBS-based beamforming. However, as the array size significantly increases, 
the conventional far-field UPW model may become invalid and the more accurate near-field spherical model needs to be taken into account. In this section, the spatial filter design and performance analysis of the CBS-based beamforming will be extended to the near-field case.

\subsection{Near-field CBS preprocessing}

To gain more useful insights, the USW model is considered for the near-field modeling, i.e., the amplitude variation across array elements is negligible \cite{lu2024tutorial}. Thus, the near-field array response vector for user $i$ is given by
\begin{equation}
	\label{array response vector for near-field}
	\left[\mathbf{a}_N \left(r_i,\theta_i\right) \right]_n
	= \frac{\sqrt{\beta_0}}{r_i} e^{-j\frac{2\pi}{\lambda}r_i}
        e^{j\frac{2\pi}{\lambda}\left(nd\sin\theta_i-\frac{n^2 d^2 \cos^2\theta_i}{2r_i}\right)},
\end{equation}
where the second-order Taylor expansion is used to approximate the phase term, i.e., $r_{i,n}\approx r_i\left(1-n\varepsilon_i \sin \theta_i+\frac12 n^2 \varepsilon_i^2\right)$. With the half-wavelength antenna spacing, we can also denote
\begin{equation}
    \label{notation for omega and psi}
    \psi_i \triangleq \frac{\pi\lambda \cos^2\theta_i}{4r_i}.
\end{equation}
The near-field array response vector \eqref{array response vector for near-field} can be simplified as 
\begin{equation}
    \label{rewritten array response vector for near-field}
    \left[\mathbf{a}_N \left(r_i,\theta_i\right) \right]_n
    =\frac{\sqrt{\beta_0}}{r_i} e^{-j\frac{2\pi}{\lambda}r_i} e^{jn\omega_i-jn^2 \psi_i},
\end{equation}
which is denoted by $a_i\left(n\right)$ for convenience.
Similarly, the resulting signal through the CBS preprocessing can be expressed as 
\begin{equation}
	\label{receive vector for near-field CBS beamforming}
	\mathbf{y}_c^{\text{near}}=\mathbf{H}_c\left(\sum_{i=1}^{K} \mathbf{a}_N \left(r_i,\theta_i\right)\sqrt{P_i} x_i + \mathbf{n}\right).
\end{equation} 


To characterize the intensity of the user $i$'s signal, we define the post-CBS vector as
\begin{equation}
    \label{near-field signal vector from user i}
    \begin{aligned}
     \mathbf{s}_i &\triangleq \mathbf{H}_c             
    \mathbf{a}_N\left(r_i,\theta_i\right)\\
        &=\begin{bmatrix}
		\sum_{l=0}^{L-1} h_c\left(l\right) a_i\left(-\widetilde{N}+L-1-l\right)\\ 
		\vdots \\ 
		\sum_{l=0}^{L-1} h_c\left(l\right) a_i\left(-\widetilde{N}+N-2-l\right)\\
		\sum_{l=0}^{L-1} h_c\left(l\right) a_i\left(-\widetilde{N}+N-1-l\right)
	\end{bmatrix}.
     \end{aligned}
\end{equation}
where $\widetilde{N} \triangleq \frac{N-1}2$.
By substituting \eqref{rewritten array response vector for near-field} into \eqref{near-field signal vector from user i},
the $p$-th element of the post-CBS vector $\mathbf{s}_i$ can be expressed as 
\begin{equation}
\label{p-th element of near-field signal vector from user i}
\begin{aligned}
    s_{i,p}
    &=\frac{\sqrt{\beta_0}}{r_i} e^{-j\frac{2\pi}{\lambda}r_i} \\
    &\quad \times \sum_{l=0}^{L-1} h_c\left(l\right) e^{j\left(-\widetilde{N}+p+L-1-l\right)\omega_i-j\left(-\widetilde{N}+p+L-1-l\right)^2 \psi_i},
\end{aligned}
\end{equation}
where $p = 0,1,\cdots,N-L$.

\subsection{Spatial filter design for near-field CBS}

For the near-field case, it is challenging to directly remove the IUI based on the frequency response of the conventional digital filter, as shown by \eqref{p-th element of near-field signal vector from user i}. Therefore, we attempt to obtain the filtering coefficients by performing the optimization based on \eqref{near-field signal vector from user i} and \eqref{p-th element of near-field signal vector from user i}. Let us first denote the filtering coefficients as a vector $\mathbf{h}=\left[h_c(0) \cdots  h_c(L-1)\right]^T$. As such, \eqref{p-th element of near-field signal vector from user i} can be further rewritten as 
\begin{equation}
    \label{inner product for p-th element of near-field signal vector}
    s_{i,p}=\frac{\sqrt{\beta_0}}{r_i} e^{-j\frac{2\pi}{\lambda}r_i}
    \mathbf{w}_{i,p}^H \mathbf{h},
\end{equation}
where $\mathbf{w}_{i,p} \in \mathbb{C}^{L \times 1}$ is a column vector and its $l$-th element, $0 \leq l \leq L-1$, can be expressed as 
\begin{equation}
	\label{l-th element of the w}
            w_{i,p,l} = e^{-j\left(-\widetilde{N}+p+L-1-     l\right)\omega_i+j\left(-\widetilde{N}+p+L-1-l\right)^2 \psi_i}.
\end{equation}

To be consistent with the analysis of the far-field case, we still use $\omega=\pi\sin\theta$ to describe the angle $\theta$ in the filter design. The set of all users' possible locations is thus defined as 
\begin{equation}
    \label{all possible locations of users}
    \mathcal{U} \triangleq \left\{\left(r,\omega\right): r \geq 0, \left|\omega\right| \leq \pi\right\}.
\end{equation}

\begin{figure}[!t]
	\centering
	\includegraphics[width=6.5cm]{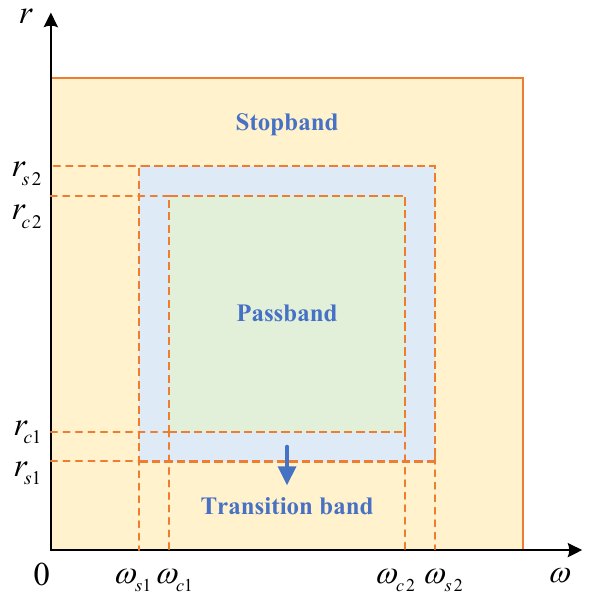}
	\caption{Near-field spatial filter design for angle and distance domain.}
	\label{near_field_filter_desig}
\end{figure}

As shown in Fig. \ref{near_field_filter_desig}, the filter's passband can be denoted by the set 
\begin{equation}
    \label{passband set}
    \mathcal{Q}_{p} \triangleq \left\{\left(r,\omega\right): r_{c1} \leq r \leq r_{c2},\omega_{c1} \leq \left|\omega\right| \leq \omega_{c2}\right\},
\end{equation}
 where $r_{c1}$, $r_{c2}$, $\omega_{c1}$ and $\omega_{c2}$ are passband cut-off values for $r$ and $\omega$, respectively. Besides, we denote the transition band as 
 \begin{equation}
    \label{transition band set}
    \mathcal{Q}_{t} \triangleq \left\{\left(r,\omega\right): r_{s1} \leq r \leq r_{s2},\omega_{s1} \leq \left|\omega\right| \leq \omega_{s2}\right\} \setminus \mathcal{Q}_{p},
\end{equation}
where $r_{s1}$, $r_{s2}$, $\omega_{s1}$ and $\omega_{s2}$ are stopband cut-off values for $r$ and $\omega$, respectively. Thus, the stopband can be expressed as
\begin{equation}
    \label{stopband set}
    \mathcal{Q}_{b} \triangleq \mathcal{U} \setminus \mathcal{Q}_p \setminus \mathcal{Q}_t.
\end{equation}

Note that we still design the angle's band in a symmetrical way, by following the conventional digital bandpass filter. For users lying in the stopband, i.e., $\mathbf{q}_i \in \mathcal{Q}_{b}$, it is expected to obtain the following result, i.e.,
\begin{equation}
	\label{anticipated gap for the interference users in the near field}
	\left\|\mathbf{s}_i\right\|  \rightarrow 0,
\end{equation}
which is equivalent to $\mathbf{s}_i \approx \mathbf{0}$. This shows that the signal intensity of all users in the set $\mathcal{Q}_{b}$ is negligible and thus these users cannot pass the spatial filter. 
Therefore, to obtain optimal filtering coefficients, the optimization problem can be formulated as 
\begin{align}
		\label{optimization problem for filter in near field}
		\min_{\mathbf{h}} &~\max_{\mathbf{q}_i \in \mathcal{Q}_{b}} \ \left\|\mathbf{s}_i \right\|,\\
		\mathrm{s.t.} &~
		\left\|\mathbf{s}_i \right\| \geq \epsilon_1, \ \mathbf{q}_i \in \mathcal{Q}_{p},\tag{\ref{optimization problem for filter in near field}a} \label{optimization problem for filter in near field a}\\
		&~\left\|\mathbf{s}_i \right\| \leq \epsilon_2, \ \mathbf{q}_i \in \mathcal{Q}_{t},\tag{\ref{optimization problem for filter in near field}b} \label{optimization problem for filter in near field b}
\end{align}
where $\epsilon_1$ and $\epsilon_2$ are thresholds for the signal intensity of passband and transition band users, respectively. Note that the objective function \eqref{optimization problem for filter in near field} aims at removing all users within the stopband set $\mathcal{Q}_{b}$. Besides, the constraint \eqref{optimization problem for filter in near field a} is to retain users in the passband and the constraint \eqref{optimization problem for filter in near field b} is to characterize the transition band of the spatial filter. However, the problem \eqref{optimization problem for filter in near field} is non-convex. By introducing a slack variable $t$, the above optimization problem is equivalent to 
\begin{align}
	\label{optimization problem for filter in near field with slack variable}
		\min_{\mathbf{h},t} &~ t,\\
		\mathrm{s.t.}
		&~\left\|\mathbf{s}_i \right\| \leq t,\ \mathbf{q}_i \in \mathcal{Q}_{b}, \tag{\ref{optimization problem for filter in near field with slack variable}a} \label{optimization problem for filter in near field with slack variable a}\\
		&~\left\|\mathbf{s}_i \right\| \geq \epsilon_1, \ \mathbf{q}_i \in \mathcal{Q}_{p},\tag{\ref{optimization problem for filter in near field with slack variable}b} \label{optimization problem for filter in near field with slack variable b}\\
		&~\left\|\mathbf{s}_i \right\| \leq \epsilon_2, \ \mathbf{q}_i \in \mathcal{Q}_{t} \tag{\ref{optimization problem for filter in near field with slack variable}c} \label{optimization problem for filter in near field with slack variable c}.
\end{align}
It can be observed that the equivalent problem \eqref{optimization problem for filter in near field with slack variable} is sill non-convex due to the constraint \eqref{optimization problem for filter in near field with slack variable b}. To address this issue, we can use the SCA technique to further relax the above optimization problem. Thus, the non-convex constraint \eqref{optimization problem for filter in near field with slack variable b} can be rewritten as 
\begin{equation}
	\label{relax form of the third constraint SCA}
	\left\|\mathbf{s}_i\right\|^2 = \frac{\beta_0}{r_i^2}\sum_{p=0}^{N-L} \mathbf{h}^H \mathbf{W}_{i,p} \mathbf{h} \geq \epsilon_1^2,
\end{equation}
where $\mathbf{W}_{i,p} \triangleq \mathbf{w}_{i,p} \mathbf{w}_{i,p}^H$. By using the first-order approximation of $\mathbf{h}^H \mathbf{W}_{i,p} \mathbf{h}$, \eqref{relax form of the third constraint SCA} can be further relaxed as 
\begin{equation}
	\label{first order in SCA technique}
	\sum_{p=0}^{N-L} \mathbf{h}_j^H \mathbf{W}_{i,p} \mathbf{h}_j + 2\mathrm{Re}\left[\mathbf{h}_j^H \mathbf{W}_{i,p} \left(\mathbf{h}-\mathbf{h}_j\right)\right] \geq \frac{\epsilon_1^2 r_i^2}{\beta_0},
\end{equation}
where $\mathbf{h}_j$ is the output of $\mathbf{h}$ at the $j$-th iteration. Thus, the primary problem \eqref{optimization problem for filter in near field with slack variable} can be transformed into \eqref{optimization problem in near field with slack variable by SCA}, shown at the top of the next page. It can be observed that the current optimization problem \eqref{optimization problem in near field with slack variable by SCA} is convex with respect to variables $\mathbf{h}$ and $t$, and can be directly solved by using CVX tools \cite{boyd2004convex}.

\begin{figure}[!t]
	\centering
	\includegraphics[width=8.4cm]{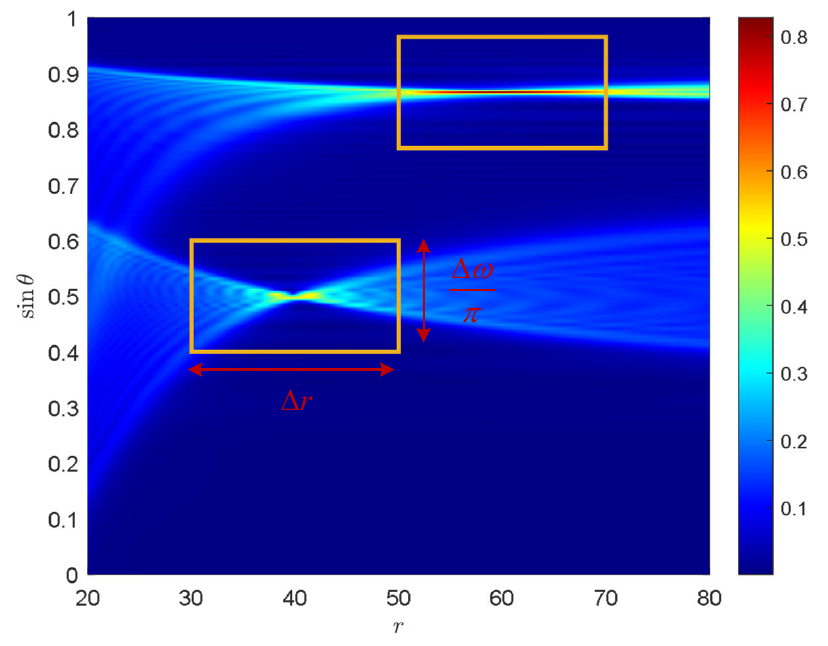}
	\caption{Correlation of near-field users, where $\sin\theta=\omega/\pi$.}
	\label{heatmap_for_near_field}
\end{figure}

\begin{figure*}[!t]
	\normalsize
	\setcounter{MYtempeqncnt}{\value{equation}}
	\setcounter{equation}{\value{MYtempeqncnt}}
	
	  
	\begin{align}
		\label{optimization problem in near field with slack variable by SCA}
			\min_{\mathbf{h},t} &~ t,\\
			\mathrm{s.t.}
			&~\left\|\mathbf{s}_i\right\| \leq t,\ \mathbf{q}_i \in \mathcal{Q}_{b}, \tag{\ref{optimization problem in near field with slack variable by SCA}a} \label{optimization problem in near field with slack variable by SCA a}\\
			&~\sum_{p=0}^{N-L} \mathbf{h}_j^H \mathbf{W}_{i,p} \mathbf{h}_j + 2\mathrm{Re}\left[\mathbf{h}_j^H \mathbf{W}_{i,p} \left(\mathbf{h}-\mathbf{h}_j\right)\right] \geq \frac{\epsilon_1^2 r_i^2}{\beta_0}, \ \mathbf{q}_i \in \mathcal{Q}_{p}, \tag{\ref{optimization problem in near field with slack variable by SCA}b} \label{optimization problem in near field with slack variable by SCA b}\\
			&~\left\|\mathbf{s}_i\right\| \leq \epsilon_2, \ \mathbf{q}_i \in \mathcal{Q}_{t} \tag{\ref{optimization problem in near field with slack variable by SCA}c} \label{optimization problem in near field with slack variable by SCA c}.
	\end{align}
	  
	\setcounter{equation}{\value{MYtempeqncnt}+1}
	\hrulefill
	\vspace*{4pt}
\end{figure*}

\subsection{Spatial filtering performance of near-field CBS}

\begin{figure*}[!t]
  \centering
  \subfloat[Filter 1 ($r_0=40 \ \text{m}$)]{\includegraphics[width=8.40cm]{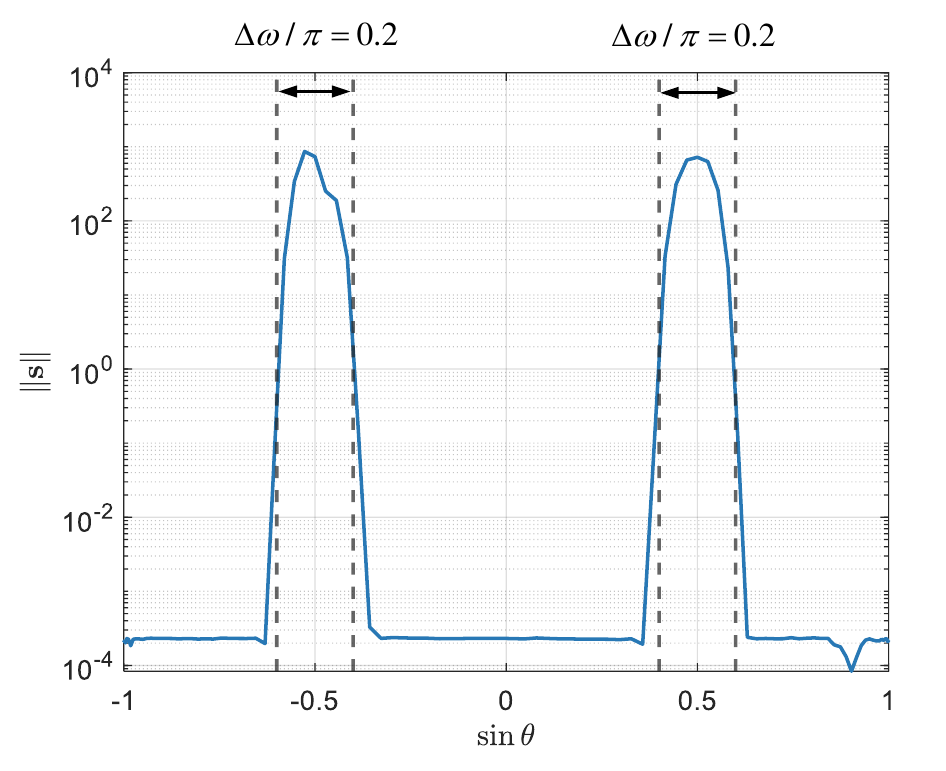}
  \label{filter_perform_r0_40}}
  \hfil
  \subfloat[Filter 2 ($r_0=60 \ \text{m}$)]{\includegraphics[width=8.40cm]{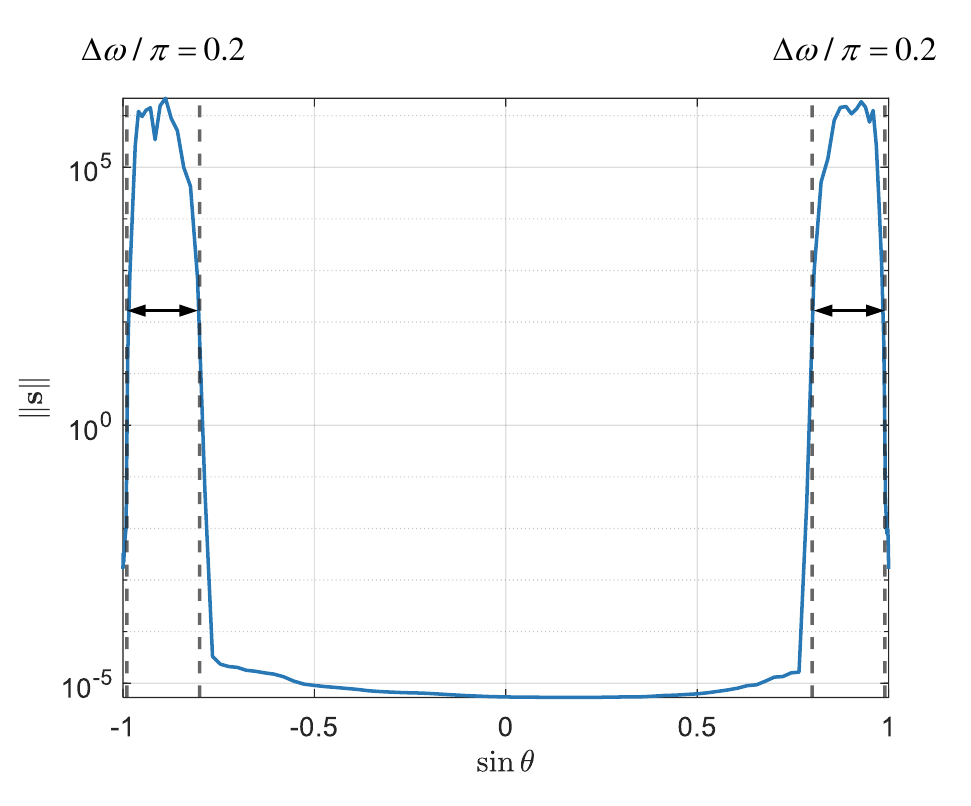}
  \label{filter_perform_r0_60}}
  \caption{Near-field spatial filters based on optimization method, where we use $\sin\theta=\omega/\pi$ to characterize its angle domain.}
  \label{Near-field filter based on optimization method}
\end{figure*}

Unlike the far-field case, near-field users will be correlated in both the angle and distance domains. To remove the IUI efficiently, we need to set the appropriate cut-off values for $r$ and $\omega$. It is worth mentioning that these cut-off values can be obtained based on the correlation of near-field users, given by
\begin{equation}
    \label{correlation of near-field users}
    \Gamma\left(r,\theta\right)\triangleq\left|\left[\mathbf{a}_N^{\text{near}}\left(r_e,\theta_e\right)\right]^H \mathbf{a}_N^{\text{near}}\left(r,\theta\right)\right|^2,
\end{equation}
where $\left(r_e,\theta_e\right)$ denotes the location of a certain user.

Fig. \ref{heatmap_for_near_field} plots the user correlation versus the distance and angle based on \eqref{array response vector for near-field} and \eqref{correlation of near-field users}, where the carrier frequency is $f=3\ \text{GHz}$ and the antenna number of the ULA is set as $N=513$. The locations of two users are assumed as $\left(r_{e,1},\theta_{e,1}\right)=\left(40\ \text{m},\pi/6\right)$ and $\left(r_{e,2},\theta_{e,2}\right)=\left(60\ \text{m},\pi/3\right)$, respectively. It can be observed that near-field users are not only correlated in the angle, but also in the distance. Besides, yellow rectangles describe the spatial filter's passband, where $\triangle r \triangleq r_{c2}-r_{c1}$ and $\triangle \omega \triangleq \omega_{c2}-\omega_{c1}$ denote the ranges of the passband for $r$ and $\omega$, respectively. Here we set $\triangle r = 20 \ \text{m}$ and $\triangle \omega =0.2\pi$. Due to the spatial filtering effect, users lying out of the passband will not be severely disturbed by those in the band, thus significantly mitigating the impact of the beam splitting in near-field multi-user communications.

To explicitly show the spatial filtering performance based on our proposed optimization method, we only consider the angle dimension first by fixing the distance as $r_0$. In this case, we have $\triangle r =0$. For user locations with $r_0$, the spatial filter's passband \eqref{passband set} reduces to 
\begin{equation}
    \label{simplied passband set}
    \mathcal{Q}_{p} =q \left\{\omega:\omega_{c1} \leq \left|\omega\right| \leq \omega_{c2}\right\}.
\end{equation}
Similarly, the transition band \eqref{transition band set} can be simplified as 
 \begin{equation}
    \label{simplified transition band set}
    \mathcal{Q}_{t} = \left\{\omega:\omega_{s1} \leq \left|\omega\right| \leq \omega_{s2}\right\} \setminus \mathcal{Q}_{p}.
\end{equation}

By substituting \eqref{stopband set}, \eqref{simplied passband set} and \eqref{simplified transition band set} into the current optimization problem \eqref{optimization problem in near field with slack variable by SCA}, we can obtain optimal filtering coefficients $\mathbf{h}$ for near-field spatial filters with different $L$, $\omega_{c1}$, $\omega_{c2}$, $\omega_{s1}$ and $\omega_{s2}$. According to the correlation shown in Fig. \ref{heatmap_for_near_field}, Table \ref{Parameters for near-field filter design} provides detailed parameters for two near-field spatial filters with $r_0=40 \ \text{m}$ and $r_0=60 \ \text{m}$, respectively. Fig. \ref{Near-field filter based on optimization method} plots the Euclidean norm of the post-CBS vector, i.e., $\left\|\mathbf{s}\right\|$, versus the angle range $\sin\theta$. It can be observed that for both filters, there exists a significant gap between the passband and stopband, rendering users lying in the stopband negligible. Besides, the amplitude of the passband is flat in general, which is nearly consistent with the amplitude-frequency response of conventional digital filters. This initially validates the feasibility of extending the CBS-based beamforming to the near-field case.

\begin{table}[!t]
	\centering
        \renewcommand\arraystretch{1.2}
        \setlength{\tabcolsep}{15pt}
	\caption{Parameters for near-field spatial filter design.} 
	\label{Parameters for near-field filter design}
\begin{tabular}{|llll|}
\hline
\multicolumn{4}{|l|}{$N=513,\epsilon_1=1,\epsilon_2=0.1$}                                                                 \\ \hline
\multicolumn{2}{|l|}{Filter 1}                                            & \multicolumn{2}{l|}{Filter 2}                       \\ \hline
\multicolumn{1}{|l|}{$\omega_{c1}$} & \multicolumn{1}{l|}{$0.4\pi$}       & \multicolumn{1}{l|}{$\omega_{c1}$} & $0.8\pi$       \\ \hline
\multicolumn{1}{|l|}{$\omega_{c2}$} & \multicolumn{1}{l|}{$0.6\pi$}       & \multicolumn{1}{l|}{$\omega_{c2}$} & $\pi$          \\ \hline
\multicolumn{1}{|l|}{$\omega_{s1}$} & \multicolumn{1}{l|}{$0.38\pi$}      & \multicolumn{1}{l|}{$\omega_{s1}$} & $0.78\pi$      \\ \hline
\multicolumn{1}{|l|}{$\omega_{s2}$} & \multicolumn{1}{l|}{$0.42\pi$}      & \multicolumn{1}{l|}{$\omega_{s2}$} & $\pi$          \\ \hline
\multicolumn{1}{|l|}{$r_0$}         & \multicolumn{1}{l|}{$40\ \text{m}$} & \multicolumn{1}{l|}{$r_0$}         & $60\ \text{m}$ \\ \hline
\multicolumn{1}{|l|}{$L$}           & \multicolumn{1}{l|}{470}            & \multicolumn{1}{l|}{$L$}           & 450            \\ \hline
\end{tabular}
\end{table}

\subsection{Near-field CBS-based beamforming}

By solving the optimization problem \eqref{optimization problem in near field with slack variable by SCA}, we are able to obtain optimal filtering coefficients $\mathbf{h}$ and corresponding Toeplitz matrix $\mathbf{H}_c$. As such, for the near-field case, the CBS-based MRC and MMSE beamforming matrices still own identical forms with \eqref{MRC matrix for CBS processing} and \eqref{modefied MMSE beamforming matrix}, respectively. However, due to the introduction of the nonlinear phase by the USW model, the linear CBS-based beamforming vector cannot be expressed as a form of the frequency response of the digital FIR filter. Specifically, for the CBS-based MRC beamforming, its linear beamforming vector for user $k$ is given by
\begin{equation}
        \label{near-field MRC vector for CBS processing}
        \mathbf{v}_{c,\text{MRC},k}^{\text{near}}=\frac{\mathbf{H}_c \mathbf{a}_{N} \left(r_k,\theta_k\right)}{\left\|\mathbf{H}_c \mathbf{a}_{N} \left(r_k,\theta_k\right)\right\|}.
\end{equation}

Similarly, for the CBS-based MMSE beamforming, its linear beamforming vector for user $k$ is
\begin{equation}
	\label{near-field CBS MMSE beamforming vector}
		\mathbf{v}_{c,\text{MMSE},k}^{\text{near}}
		= \frac{\mathbf{C}_{c,k}^{-1} \mathbf{H}_c \mathbf{a}_{N} \left(r_k,\theta_k\right)}{\left\|\mathbf{C}_{c,k}^{-1} \mathbf{H}_c \mathbf{a}_{N} \left(r_k,\theta_k\right)\right\|},
\end{equation}
where $\mathbf{C}_{c,k}^{-1}$ is still determined by \eqref{interference-plus-noise covariance matrix}.

\section{Numerical results}

In this section, numerical results are provided to validate our proposed CBS-based beamforming, both in the far-field and near-field cases. Unless otherwise stated, we assume the identical transmit SNR $\bar{P}$ for each user. Besides, the one-ring model is utilized to deploy scatterers in the multi-path environment.

\subsection{Far-field case}

\begin{figure}[!t]
	\centering
	\includegraphics[width=7.7cm]{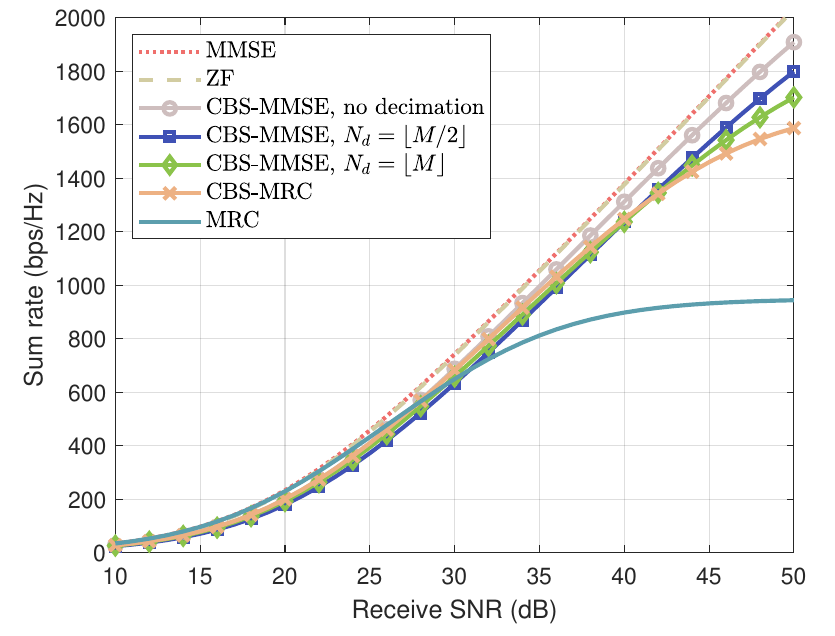}
	\caption{Sum rate versus receive SNR.}
	\label{CBS_sum_rate}
\end{figure}

Fig. \ref{CBS_sum_rate} plots the sum rates versus the receive SNR $\beta_0 \bar{P}$, where the antenna number at the BS is set as $N=513$ and $K=200$ users are simultaneously served. We assume that all users are uniformly distributed over the angular frequency domain $\omega_i \in \left[-\pi,\pi\right]$ with the equal distance $r_i=200\ \text{m}$. Besides, each user is corresponding to the one-ring model with $3$ randomly deployed scatterers, whose radius is set as $5\ \text{m}$.
We compare the resulting sum rates based on the CBS-based MRC beamforming \eqref{MRC matrix for CBS processing}, the CBS-based MMSE beamforming \eqref{modefied MMSE beamforming matrix}, the CBS-based MMSE beamforming using the array decimation \eqref{CBS MMSE beamforming vector after decimation} with three classical linear beamforming schemes introduced in Section \uppercase\expandafter{\romannumeral3}. For the spatial filter design, the length of the bandpass FIR filter is set as $L=110$ with its passband width being $\triangle \omega = \frac{\pi}9$. As such, the corresponding filter bank is composed of $M=9$ filters with different cut-off frequencies, i.e., $\omega_{c1}^m = (m-1)\triangle \omega$ and $\omega_{c2}^m = m\triangle \omega$, where $m=1,2,\cdots,M$ is the filter index. It can be shown that our proposed CBS-based beamforming schemes achieve a similar performance with classical ones at the low receive SNR. As the receive SNR increases, the slight performance loss can be observed for the CBS-based MRC beamforming and CBS-based MMSE beamforming using the array decimation, but there still exists a significant gap between the proposed CBS-based beamforming and the classical MRC. Furthermore, the CBS-based MMSE beamforming with no decimation obtains a near-optimal performance of the IUI mitigation, compared with the conventional MMSE and ZF. This demonstrates the effectiveness of performing the CBS preprocessing in MU-MIMO communications .

\begin{figure}[!t]
	\centering
	\includegraphics[width=7.7cm]{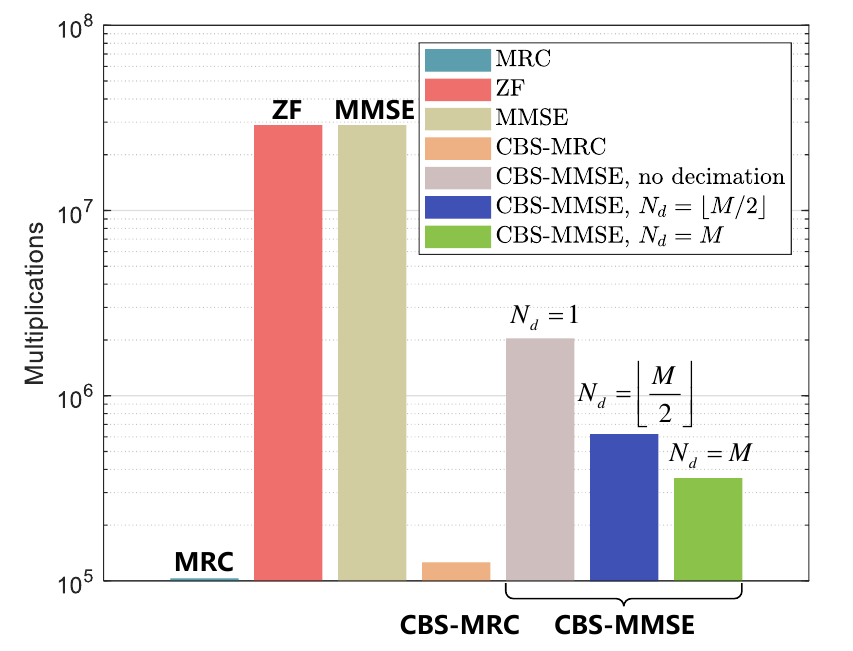}
	\caption{Complexity of different linear beamforming schemes.}
	\label{complexity_analysis}
\end{figure}

Fig. \ref{complexity_analysis} compares the complexity of our proposed CBS-based beamforming with classical linear schemes, based on TABLE \ref{Complexity for MRC, ZF, MMSE and CBS} and TABLE \ref{Complexity for CBS-MMSE beamforming}, respectively. Note that the corresponding simulation parameters are similar with those used in Fig. \ref{CBS_sum_rate}. It can be observed that under communication scenarios with massive user access, there is a considerable computational cost for classical linear schemes like the MMSE and ZF. However, our proposed CBS-based MMSE beamforming can significantly reduce the complexity at nearly the same performance level. Moreover, by using the array decimation, the corresponding complexity can be further lowered as expected, despite an acceptable performance loss. Besides, it can be shown that the CBS-based MRC beamforming and classical MRC own similar levels of the computational complexity, but the CBS-based MRC beamforming obviously outperforms the classical one and meanwhile approaches the performance of the optimal scheme. Thus, the CBS-based MRC beamforming acheieves a better trade-off between the complexity and performance.

\begin{figure}[!t]
	\centering
	\includegraphics[width=7.5cm]{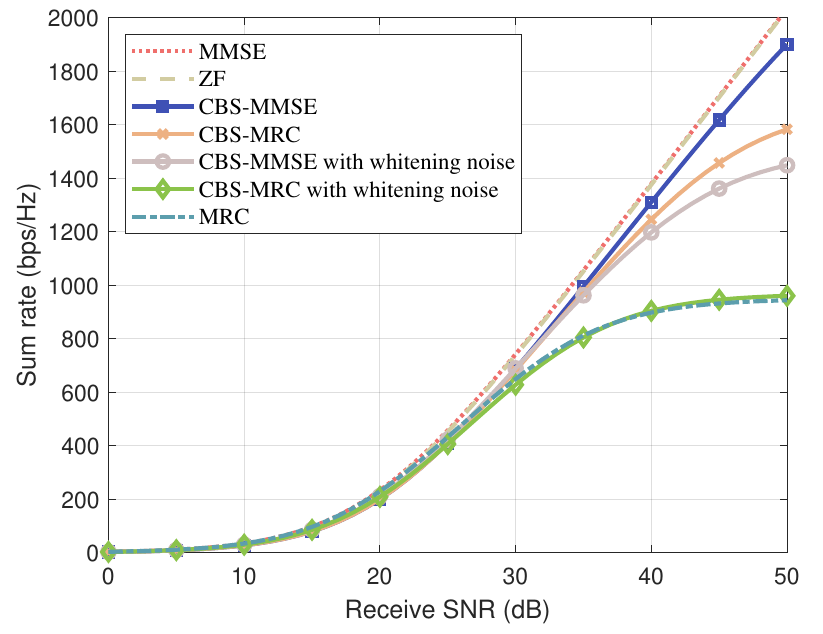}
	\caption{Sum rate versus receive SNR by considering noise whitening.}
	\label{CBS_whitening_nois}
\end{figure}

Fig. \ref{CBS_whitening_nois} shows the impact of the noise whitening on the CBS-based beamforming, based on the parameter configuration used in Fig. \ref{CBS_sum_rate}. For both the CBS-based MRC and CBS-based MMSE beamforming, significant performance degradation can be noticed. In particular, the performance of the CBS-based MRC beamforming closely aligns with that of the classical MRC. This can be explained by the nonlinear power gain, which is introduced by the whitening operator $\mathbf{W}_c$. Although it can whiten the non-white noise, the IUI will also be enlarged by multiplying with the whitening operator, thus resulting in the unsatisfactory outcomes. Moreover, by taking into account the additional computational complexity brought by the whitening operator, it unveils that for the CBS-based beamforming, the impact of the non-white noise can be neglected in the practical CBS-based beamforming design.

\subsection{Near-field case}

\begin{figure}[!t]
	\centering
	\includegraphics[width=7.5cm]{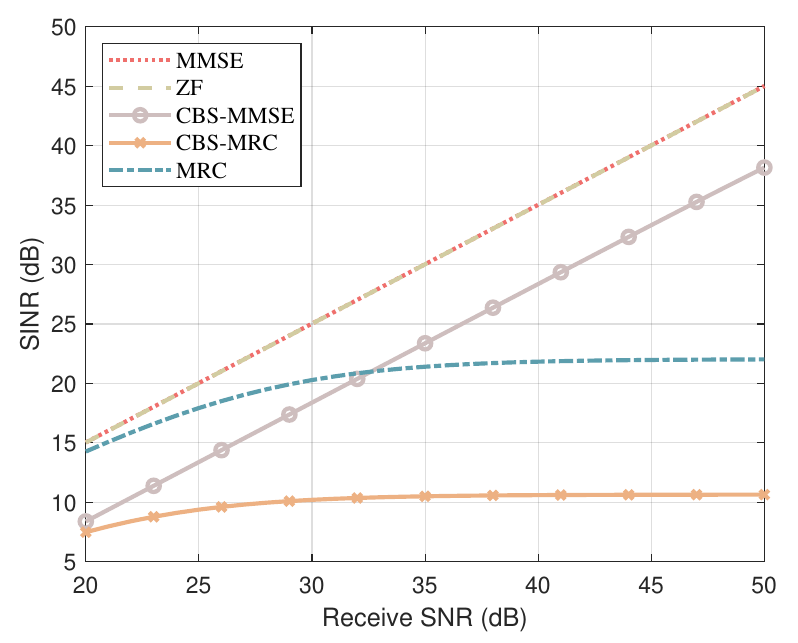}
	\caption{SINR of user $i$ versus receive SNR in the near-field case.}
	\label{CBS_near_field_1}
\end{figure}

Fig. \ref{CBS_near_field_1} plots the SINR of a certain user located in the passband versus the receive SNR $\beta_0 \bar{P}$, where the antenna number at the BS is set as $N=513$ and there are $K=100$ users that the BS simultaneously serves. Similarly, we assume that all users are uniformly distributed over the angular domain $\theta_i \in \left[-\pi/2,\pi/2\right]$ with the equal distance $r_i=40\ \text{m}$. Then, each user is also corresponding to the one-ring model with $3$ randomly deployed scatterers, whose radius is set as $5\ \text{m}$. Note that numerical results of the CBS-based beamforming  are mainly obtained based on \eqref{near-field MRC vector for CBS processing} and \eqref{near-field CBS MMSE beamforming vector}. In terms of the near-field spatial filter design, we set the filter length as $L=420$ with the passband being $\left[0.6\pi,0.8\pi\right]$. It can be observed that compared with the classical MRC, the performance of the CBS-based MRC beamforming becomes worse. This is because filtering coefficients $\mathbf{h}$ based on the optimization problem \eqref{optimization problem in near field with slack variable by SCA} significantly amplify both the IUI and noise. Besides, for the CBS-based MMSE beamforming, it can achieve an excellent performance at the high receive SNR since the enlarged IUI can be significantly mitigated by performing the MMSE on passband users. However, due to the enlarged noise, there still exists an about 7-dB gap by comparison with the optimal MMSE scheme.

\section{Conclusions}

This paper proposed a low-complexity CBS-based beamforming scheme for the IUI mitigation in MU-MIMO communications. 
The CBS-based MRC beamforming was firstly studied base on the far-field UPW model, including its beamforming design and complexity analysis. We also gained more insights into the CBS-based MMSE beamforming, where the complexity can be further reduced via the array decimation. In particular, trade-offs between the performance and complexity were clearly revealed for these two schemes. Furthermore, we extended the CBS-based beamforming to the near-field case by using a novel optimization-based CBS approach. Numerical results validated our theoretical analysis and demonstrated the effectiveness of our proposed beamforming solution.




\ifCLASSOPTIONcaptionsoff
  \newpage
\fi



%



\bibliographystyle{IEEEtran}
\bibliography{bare_jrnl}

%








\end{document}